\documentclass[12pt]{article}
\usepackage[utf8]{inputenc}
\usepackage{graphicx}
\usepackage{amsmath}
\usepackage{amsthm}
\usepackage{amssymb}
\usepackage{textcomp}
\usepackage{siunitx}
\usepackage{subfigure}
\usepackage{titling}
\usepackage{xcolor}
\usepackage{afterpage}
\usepackage{algorithm} 
\usepackage{algpseudocode} 
\usepackage{multirow}
\usepackage{hyperref}
\usepackage{algpseudocode}
\usepackage{mathtools}
\hypersetup{
    colorlinks=true,
    linkcolor=blue,
    citecolor = brown,
    filecolor=magenta,      
    urlcolor=cyan,
    pdftitle={Overleaf Example},
    pdfpagemode=FullScreen,
    }
\usepackage[font=small,labelfont=bf]{caption} 
\usepackage{graphicx} 
\usepackage{booktabs}
\usepackage{geometry}
\newtheorem{definition}{Definition}

\newcommand{\N}{\mathbb{N}}
\newcommand{\E}{\mathbb{E}}

\newcommand{\argmax}{\text{argmax}}

\newcommand{\sign}{\textrm{sign}}

\title{Online Learning of Order Flow and Market Impact\\ with Bayesian Change-Point Detection Methods}

\author{Ioanna-Yvonni Tsaknaki\thanks{Scuola Normale Superiore, Pisa, Italy. Email address: ioannayvonni.tsaknaki@sns.it} \and Fabrizio~Lillo\thanks{Dipartimento di Matematica, Universit\`a di Bologna and Scuola Normale Superiore, Pisa, Italy. Email address: fabrizio.lillo@unibo.it} \and Piero Mazzarisi\thanks{Universit\`a di Siena, Siena, Italy. Email address: piero.mazzarisi@unisi.it}}

\date{}

\begin{document}

\maketitle

\begin{abstract}
Financial order flow exhibits a remarkable level of persistence, wherein buy (sell) trades are often followed by subsequent buy (sell) trades over extended periods. This persistence can be attributed to the division and gradual execution of large orders. Consequently, distinct order flow regimes might emerge, which can be identified through suitable time series models applied to market data. In this paper, we propose the use of Bayesian online change-point detection (BOCPD) methods to identify regime shifts in real-time and enable online predictions of order flow and market impact. To enhance the effectiveness of our approach, we have developed a novel BOCPD method using a score-driven approach. This method accommodates temporal correlations and time-varying parameters within each regime. Through empirical application to NASDAQ data, we have found that: (i) Our newly proposed model demonstrates superior out-of-sample predictive performance compared to existing models that assume i.i.d. behavior within each regime; (ii) When examining the residuals, our model demonstrates good specification in terms of both distributional assumptions and temporal correlations; (iii) Within a given regime, the price dynamics exhibit a concave relationship with respect to time and volume, mirroring the characteristics of actual large orders; (iv) By incorporating regime information, our model produces more accurate online predictions of order flow and market impact compared to models that do not consider regimes.
\end{abstract}

\section{Introduction}

The study and modeling of order flow and market impact in financial markets hold paramount importance for understanding the incorporation of private information into prices and designing effective trading algorithms that consider transaction costs. A substantial body of literature (see for example \cite{em:09,Lillo23} and references therein) has revealed that the joint modeling of impact and order flow is more intricate than initially presumed.

The persistence and autocorrelation of signed trade order flow\footnote{I.e. the sequence of signed trade volume, positive (negative) when buyer (seller) initiated.} have been extensively documented since the works of \cite{LilloFarmer04} and \cite{Bouchaud04}. This persistence aligns with a long memory process, suggesting that a realistic market impact model should combine statistically efficient prices with correlated order flow. The introduction of transient impact models, also known as propagator models (\cite{Bouchaud04}), successfully accomplishes this goal. Additionally, empirical evidence has attributed the temporal persistence of order flow primarily to order splitting, as discussed in \cite{Toth}. Order splitting refers to the common practice of large investors incrementally executing their orders, termed ‘‘metaorders," through several smaller trades known as ‘‘child orders." The model proposed by \cite{ar:2005} quantitatively establishes a relationship between the autocorrelation of order flow and the distribution of metaorder sizes. In essence, this model postulates that the strong serial dependence arises from the optimal execution strategies employed by institutional investors, which leads to persistent order flow.

From an econometric perspective, this notion might be connected to the well-known fact (see \cite{GrangerHyung,MikoschStarica,r:01}) that long memory time series can be (approximately) generated by regime-shift models, where each regime exhibits short memory and heterogeneous lengths. Regime shift models gained popularity around two decades ago when they were proposed to explain the long-range memory of volatility, as seen in \cite{r:01}.

This paper proposes to use regime shift models to describe order flow time series, with the objectives of: (i) econometrically explaining the long memory of order flow; (ii) enhancing the prediction of order flow and price dynamics through detected regimes; and (iii) suggesting a connection between regimes and the execution of metaorders. Unlike many conventional regime shift approaches that require a predetermined number of regimes (e.g., Hidden Markov Models), we focus on a model that allows online detection of change-points (CPs) to identify the occurrence of new regimes. Existing algorithms typically operate offline, recursively segmenting the time series ex-post into increasingly smaller regimes.

In this paper, we specifically concentrate on online CP detection, employing Bayesian approaches. The Bayesian framework is well-suited for quantifying our uncertainty regarding CPs using the posterior distribution, as illustrated in \cite{r:15}. We primarily consider the class of algorithms known as Bayesian Change-Point Detection Methods (BOCPD), pioneered by \cite{c:07} as an improvement on the ideas developed by \cite{r:07}. Since 2007, BOCPD and its extensions have found applications in various financial settings. Most applications have focused on stock returns, as demonstrated in \cite{r:2017}, \cite{c:2020}, \cite{c:2022}, and more recently, \cite{r:22,LleoZiembaLi} utilized BOCPD as an exit-entry model for long-short prediction in the stock market. The work by \cite{r:2017} extended the BOCPD approach to the multi-sequence setting to analyze changes in 401 U.S. stocks within the S\&P 500 index. BOCPD utilizes a message-passing algorithm to recursively compute the posterior distribution of the time since the last CP, termed the ‘‘run length". This elapsed time is continuously updated upon receiving new data points. To perform online inference, the underlying predictive model (UPM) is computed, representing the distribution of data given the current run length. For instance, the UPM may assume a Gaussian model with different means across regimes.

In the BOCPD model, the data is assumed to be independently and identically distributed (i.i.d.) within each regime. However, this assumption is unrealistic for most financial time series. In this work, we extend BOCPD to accommodate Markovian data within each regime. While \cite{c:07b} consider the correlation structure of multivariate time series, their CP detection algorithm is offline. To the best of our knowledge, this is the first work that combines an online learning algorithm for CPs with a Markovian data structure. Furthermore, we propose a second extension of the BOCPD algorithm that relaxes the assumption of constant parameters within a regime, allowing for time-varying autocorrelation. To achieve this, we employ the class of Score Driven models introduced by \cite{Score-Driven1} and \cite{Harvey}, which provide an observation-driven framework for real-time learning of time-varying parameters. Thus, our newly proposed method combines the online CP detection approach of BOCPD with the online learning of time-varying autocorrelation parameters within each regime.

We present an empirical application of the proposed methods using order flow samples from stocks traded on NASDAQ. In an out-of-sample forecasting exercise, we find that the Score-Driven-based method outperforms other models, including autocorrelated time series models without regimes. Our analysis demonstrates that the model is correctly specified, and the residuals within each regime exhibit no correlation. By investigating the price dynamics during identified regimes, we discover that they follow concave functions of time, with the total price change in a regime also exhibiting a concave relationship with volume. These findings resemble those observed for real metaorders, consistent with the square root impact law. Finally, we demonstrate that knowledge of order flow regimes can be effectively utilized to improve predictions of order flow and price dynamics. We accomplish this by exploiting the well-known correlation between order flow and simultaneous/future price changes through market impact.

The paper is organized as follows: In Section 2, we present the dataset, the variable of interest, and provide the motivation for applying a regime shift model to order flow time series. Section 3 covers the methodological aspects of the paper, outlining the main properties of BOCPD and introducing the two proposed extensions. In Section 4, we describe the estimation results of the models on NASDAQ data and analyze the obtained findings. We examine the average price dynamics during an order flow regime and quantify the relationship between the total price change and the net order flow exchanged within a regime. Moreover, it is presented the forecasting analysis of order flow and the correlation with price dynamics through market impact. Finally, in Section 5, we draw conclusions and offer suggestions for further research.
\section{Data Set and Motivation}

In this paper, we consider the order flow of trades and in particular the aggregated signed volume. Let $M$ be the number of trades in a given day and let $v_i$  ($i=1,\ldots,M$) the signed volume (positive for buyer initiated  and negative for seller initiated trades) of the $i$-th trade and let us indicate with $N$ the number of trades we aggregate, so that our time series is composed by $T=\lfloor M/N \rfloor$ observations. The time series of interest is the aggregated order flow $x_t$ on the {\it interval}  $~\N\cap[N(t-1)+1,N(t-1)+N]$ given by 
\begin{equation}
    x_t = \sum_{j=1}^N v_{N(t-1)+j},~~~~~~~~~~t=1,...,T.
\end{equation}

Our data set consists of executed orders during March 2020 for Microsoft Corp. (MSFT) and of December 2021 for Tesla Inc. (TSLA).
In order to investigate the role of aggregation time scale, we choose $N=240$ and $N=730$ for TSLA  and 
$N=400$ and $N=1200$ for MSFT, which, in both cases correspond to an average time interval of 1 and 3 minutes, respectively.
The length of the two time series is 8,686 data points for 1 minute and 2,856 data points for 3 minutes of TSLA and 8,723 data points for 1 minute and 2,908 data points for 3 minutes of MSFT.
The choice of 1 and 3 minutes is of course arbitrary. However we decided to choose these interval lengths in order to avoid any microstructure noise but being still at high frequency.

\begin{figure}
    \centering
    \subfigure[]{\includegraphics[width=0.43\textwidth]{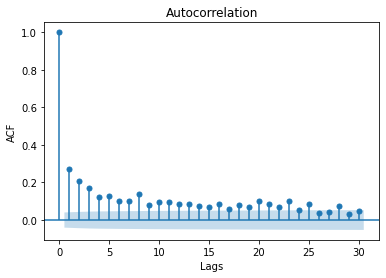}} 
    \subfigure[]{\includegraphics[width=0.43\textwidth]{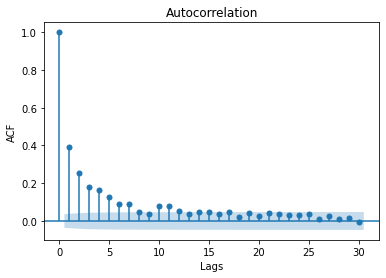}}
    \caption{Autocorrelation function of the order flow aggregated at the 3 minute time scale corresponding to $N=730$ executions for TSLA (a), and $N=1200$ executions for MSFT (b).}
    \label{fig1}
\end{figure}

To motivate our analysis, Figure \ref{fig1} shows the autocorrelation function of the order flow for TSLA and MSFT at the 3 minutes aggregation scale. Consistently with the literature (see, e.g., \cite{LilloFarmer04,book:18}), we observe that the autocorrelation function in all the cases has a slow decay. i.e. a buy (sell) trade is more likely followed by a buy (sell) trade. More quantitatively, it has been documented in many papers (see \cite{Bouchaud04,LilloFarmer04,BouchaudKockelkorenPotters,em:09,YamamotoLebaron,Toth,Taranto-TothA,Taranto-TothB}) that the autocorrelation function of trade signs $\rho(\tau)$ decays asymptotically as a power law with an exponent smaller than one
$$
\rho(\tau)\sim \frac{1}{\tau^\gamma}
$$
where $\gamma<1$ which implies that the time series is long memory with a Hurst exponent $H=1-\gamma/2>1/2$. 

The origin of this large persistence has been investigated both empirically and theoretically. Making use of labeled data allowing to identify the market member initiating each trade, \cite{Toth} empirically showed that the long-range persistence observed at the London Stock Exchange is strongly driven by order splitting, i.e. the same trader sequentially placing trades with the same sign, very likely as part of an optimal execution program. On the contrary, herding, i.e. groups of investors trading in the same direction in the same period, plays a much minor role. 

From a theoretical point of view, the connection between order splitting and long memory of order flow has been elucidated by \cite{ar:2005} (LMF). They proposed a simple model that postulates that market participants who intend to execute large orders split them into smaller orders and trade them incrementally. The large orders are called metaorders and the small trades in which they are split and sequentially traded are termed child orders. Under the assumption that metaorders are randomly sampled from a size distribution $p_L$ ($L\in {\mathbb N})$, the LMF model predicts the form of the autocorrelation function of trade signs. In particular, when the distribution $p_L$ is a power law
\begin{equation}
    p_L = \frac{\alpha}{L^{1+\alpha}}
\end{equation}
then the autocorrelation function of trade signs decays asymptotically as
\begin{equation}
    \rho(\tau)\sim \frac{1}{\tau^{\alpha-1}}
\end{equation}
i.e. the model predicts that $\gamma=1-\alpha$. Interestingly, very recently \cite{kana23} theoretically showed that predictions of the LMF model remain valid also when there is heterogeneity in trading frequency and size distribution across market participants.

The empirical validation of the LMF model poses some challenges because of the need for complete information on metaorders traded in the market. \cite{ar:2005} used off-market trades as a proxy of metaorders. An alternative to such proxy is suggested by  \cite{r:20,r:09,r:80}, who proposed segmentation algorithms and Hidden Markov Models to identify  metaorders from brokerage data. Without relying on noisy proxies, \cite{ar:2013,r:2015}  used private data of real metaorders by financial companies to test the power law hypothesis. In such a case, the results lacks generality since information on metaorders is company-specified. Recently, \cite{c:23} used account-level data of the whole Tokyo Stock Exchange to directly test the LMF model. The predicted relationship between the exponent $\gamma$ and $\alpha$ has been very accurately verified, both at the market and at the single stock level.

Summarizing, the LMF model proposes that most of the autocorrelation of order flow comes from the execution of metaorders but, due to the anonymous nature of financial markets, their presence cannot be easily and directly inferred  from public market data. However, the start of the execution of a new metaorder should lead to a regime change in the order flow time series, which could be detected with suitable statistical methods. Thus the identification of a CP in the order flow might signal the arrival of a new metaorder execution. Moreover, from the market (or econometric) point of view, the identification of a CP modifies the forecasting of future order flow and price dynamics, since only past data after the last CP are useful for predictions. The practical use of such methods for CP detection requires that they work {\it online}, i.e. that the identification is done in real time and not {\it ex-post} (such as in the segmentation algorithms used, for example, in \cite{r:20}). Finally, the method should allow to perform one-step ahead prediction in an online mode.

\section{Bayesian Online Change-Point Detection Algorithms}\label{sec: Bayesian Online Change-Point Detection Algorithms}

In the following, we briefly review the BOCPD algorithm, introduced by \cite{c:07}, and present two novel extensions that we propose here, namely the Markovian BOCPD (MBO) and the Markovian BOCPD for Correlated data (MBOC) algorithm. The original BOCPD algorithm relies on the assumption that the data are independent and identically distributed within each regime. To relax such a strong assumption, we propose a new MBO algorithm that considers Markovian dynamics within each regime, thus allowing for serial correlation. At this stage, both BOCPD and MBO assume that parameters (i.e. mean, variance, autocorrelation) are constant within each regime. We further relax this assumption in a generalization of the MBO, named MBOC, that accounts for time-varying correlations. Such a generalization is based on the Score Driven approach introduced by \cite{Score-Driven1}. In a companion paper \cite{tsaknaki}, we introduce in full generality the new class of regime-shift score-driven models, where other parameters can change over time within each regime.

\subsection{The BOCPD Algorithm}\label{subsec:BOCPD}
The BOCPD algorithm for i.i.d. data has been introduced by \cite{c:07}.
Let $x_{1:T} = \{x_1,...,x_T\}$ be a sample time series.  The model assumes that data are non-stationary (because of regimes) and satisfy the product partition model (PPM), see \cite{r:92}, meaning that data can be partitioned into regimes. Moreover, the parameters $\theta_{R}$ within each regime $R$ are i.i.d. random variables drawn from some given distribution. Such a distribution needs to belong to the exponential family. Throughout the paper, we consider normal distributions. This assumption is tested using a Jarque-Bera (JB) statistic in the empirical application as shown in Section \ref{sec: Online Prediction of Order Flow}. 

\cite{c:07}, assume a time series as the realization of i.i.d. random variables from a normal distribution with unknown mean $\theta_{R}$ and known variance $\sigma^2$,
\begin{equation}\label{eq4}
    x_i\sim\mathcal{N}(\theta_{R},\sigma^2).
\end{equation}

Regimes and CPs separating them are not directly observable but must be inferred from data. To this end, the goal is to infer the elapsed time since the last CP, a quantity named {\it run length} and defined as follows.
\begin{definition}
    The run length $r_t$ is a non-negative discrete variable defined as:
    \begin{equation}
    r_t = 
    \begin{cases}        0,\quad\quad\quad\quad\text{if a CP occurs at time $t$}\\
        r_{t-1}+1,\quad\text{else}.
    \end{cases}
\end{equation}
\end{definition}
In the BOCPD algorithm, the arrival of a CP is modeled as a Bernoulli process\footnote{Other assumptions on the distribution of regimes can be implemented, for example, those leading to a non-exponential distribution of regime length. This more realistic extension is left for future research.} with hazard rate $1/h$:
\begin{equation}
    p(r_t|r_{t-1}) = \begin{cases}
        1/h,\quad\quad\quad\text{if $r_t=0$}\\
        1-1/h,\quad\text{if $r_t = r_{t-1}+1$}\\
        0,\quad\quad\quad\quad\text{otherwise}.
    \end{cases}
\end{equation}
The primary quantity of interest is the computation of the run length posterior $p(r_t|x_{1:t})$ which characterizes probabilistically the number of time steps since the last CP given the data observed so far,
\begin{equation}\label{eq:RL posterior}
    p(r_t|x_{1:t}) = \frac{p(r_t,x_{1:t})}{p(x_{1:t})}.
\end{equation}
The joint distribution over both the run length and the observed data can be written recursively,
\begin{align}
    p(r_t,x_{1:t}) & = \sum_{r_{t-1}}p(r_t,r_{t-1},x_t,x_{1:t-1})\\
    \label{eq:joint distr 2} & =
        \sum_{r_{t-1}}\underbrace{p(x_t|r_{t-1},x_{1:t-1})}_\text{UPM}\underbrace{p(r_t|r_{t-1})}_\text{Hazard}\underbrace{p(r_{t-1},x_{1:t-1})}_\text{Message}.
\end{align}
An important assumption that simplifies the computation is about the {\it changepoint prior}, namely $r_t$ is conditionally dependent on $r_{t-1}$ only.
The quantity $p(x_{1:t})$  is named {\it evidence} and is computed as
\begin{equation}\label{eq:evidence}
    p(x_{1:t}) = \sum_{r_t}p(r_t,x_{1:t}).
\end{equation}
The Underlying Predictive Model (UPM) is defined as the predictive posterior distribution given the current run length. Because of the assumption on PPM, such a distribution depends only on the last $r_{t-1}$ observations and can be stated in a more compact form as
\begin{align}
    p(x_t|r_{t-1},x_{1:t-1}) & = p(x_t|x_{t-1}^{(r_{t-1})})
\end{align}
where
\begin{equation}
    x_{t-1}^{(r_{t-1})} = x_{t-r_{t-1}:t-1}\quad\text{and}\quad x_{t:t-1} = \emptyset .
\end{equation}
By using the conjugacy property of the exponential family when data are i.i.d., one obtains closed-form solutions for the UPM term, see \cite{r:79, k:07}. For normal distributions, the UPM term is
\begin{align}\label{eq13}
    p(x_{t}|x_{t-1}^{(r_{t-1})}) & = \mathcal{N}(\mu_{r_{t-1}},\sigma^2+\sigma_{r_{t-1}}^2),
\end{align}
with posterior parameters given by
\begin{equation}\label{eq: post params BOCPD}
    \mu_{r_{t-1}} = \frac{\frac{\sum_{i=t-r_{t-1}}^{t-1}x_i}{\sigma^2}+\frac{\mu_0}{\sigma_0^2}}{\frac{r_{t-1}}{\sigma^2}+\frac{1}{\sigma_0^2}}\quad\text{and}\quad \sigma_{r_{t-1}}^2 = \Big(\frac{r_{t-1}}{\sigma^2}+\frac{1}{\sigma_0^2}\Big)^{-1}\quad\text{for}\quad r_{t-1}\in\{1,...,t-1\}.
\end{equation}

Let us stress that the run length in Eq. (\ref{eq: post params BOCPD}) is a latent variable we must infer. As such, the value of the posterior parameters varies depending on $r_{t-1}$, i.e. where we put the last change point, whose probability is in Eq. (\ref{eq:RL posterior}).

The BOCPD algorithm works as follows. At time $t=0$, we initialize the prior values $\mu_0, \sigma_0^2$ and the known variance $\sigma^2$ (see below for details). At \textcolor{blue}{a} generic time $t>0$, a new data point $x_t$ becomes available, and the UPM in Eq.~(\ref{eq13}) is computed for any possible $\mu_{r_{t-1}}$ and $\sigma^2_{r_{t-1}}$, as a function of the run length $r_{t-1}$ that takes value from $0$ to $t-1$. Then, the joint distribution over both the run length and the observed data point, see Eq. (\ref{eq:joint distr 2}), is computed for all the possible values of the run length. Thus we obtain:
\begin{enumerate}
    \item the growth probabilities,
    \begin{equation}
       p(r_t=l,x_{1:t}),\quad\text{for } l = 1,...,t;
    \end{equation}
    \item the CP probability
    \begin{equation}
        p(r_t=0,x_{1:t}).
    \end{equation}
\end{enumerate}
After computing the evidence in Eq. (\ref{eq:evidence}), the run length posterior is obtained by Eq.~(\ref{eq:RL posterior}). Finally, $\mu_{r_t}$ and $\sigma^2_{r_t}$ are updated as in Eq. (\ref{eq: post params BOCPD}) in order to be used at the next time $t+1$.

\subsection{The MBO Algorithm}
The assumption about the independence of data is clearly restrictive. Here we introduce the MBO algorithm as an extension of the BOCPD algorithm to the case of Markovian dependence. Similarly to before, we consider normally distributed data. As such, within a regime $R$, a time series is a realization of an AR(1) process with normal innovations,
\begin{align}\label{eq15}
    x_t & \sim \mathcal{N}(\theta_{R},\sigma^2),\\\label{eq16}
    x_t|x_{t-1} & \sim \mathcal{N}(\theta_{R}+\rho(x_{t-1}-\theta_{R}),\sigma^2(1-\rho^2)).
\end{align}
As in the previous model, in each regime, the unconditional distribution is normal with unknown mean $\theta_{R}$ and known variance $\sigma^2$. Moreover the conditional distribution is normal with constant correlation $\rho = \frac{\text{Cov}(x_t,x_{t-1})}{\sigma^2}$. In the next Section, we introduce a further generalization, allowing $\rho$ to be time-varying within each regime\footnote{ We have also explored a further generalization where the variance inside each regime is time-varying, similarly to a GARCH model. The results for the order flow are qualitatively similar and we do not present them here. The reader interested in this model can consult the companion paper \cite{tsaknaki}.}.

The key observation is that the conjugacy property still holds when the data are Markovian since the conditional distribution of any member in the exponential family is still in the family, see \cite{r:08}. After some computations, one can obtain a closed form for the UPM term as
\begin{align}
    p(x_{t+1}|x_t^{(r_t)})
    & = \mathcal{N}\big(\mu_{r_t}+\rho(x_t-\mu_{r_t}),\sigma^2(1-\rho^2)+\sigma_{r_t}^2(1-\rho)^2\big),
\end{align}
where the posterior parameters are
\begin{align}\label{eq: post params MBO}
    \mu_{r_t} = \frac{b_{r_t}+\frac{\mu_0}{\sigma_0^2}}{a_{r_t}+\frac{1}{\sigma_0^2}}\quad\text{and} \quad\sigma_{r_t}^2 = \Big(a_{r_t}+\frac{1}{\sigma_0^2}\Big)^{-1}\quad\text{for}\quad r_t\in\{1,...,t\}
\end{align}
and
\begin{align}
    a_{r_t} & = \frac{1}{\sigma^2}+\frac{(r_t-1)(1-\rho)^2}{\sigma^2(1-\rho^2)}\\
    b_{r_t} & = \begin{cases}\frac{x_{t}}{\sigma^2},\quad\text{for}\quad r_t=1\\[10pt] 
    \frac{x_{t-1}}{\sigma^2}+\frac{(1-\rho)(x_t-\rho x_{t-1})}{\sigma^2(1-\rho^2)},\quad\text{for}\quad r_t=2\\[10pt]
        \frac{x_{t+1-r_t}}{\sigma^2}+\frac{(1-\rho)^2\sum_{i=t+2-r_t}^{t-1}x_i+(1-\rho)(x_t-\rho x_{t+1-r_t})}{\sigma^2(1-\rho^2)},\quad\text{for}\quad r_t\in\{3,...,t\}
    \end{cases}.
\end{align}
Let us notice that we recover the previous case when $\rho=0$.

\subsection{The MBOC Algorithm}
\begin{algorithm}[tb]
\caption{MBOC}
\label{alg: MBOC}
\textbf{Input}: $\mu_0,\sigma_0^2,\vec\lambda, \rho_1, \sigma_i^2, p(r_0=0) = 1$, $\eta$\\
\textbf{Output}: $p(r_t|x_{1:t}), \hat{\mu}_t$
\begin{algorithmic}[1] 
\For{$t = 1,...$} 
\State Observe $x_t\sim\mathcal{N}(\theta_{R},\sigma^2)$
\State Compute $\hat{\mu}_t$
\State Find $\text{argmax}_{i\in\{0,1,...,t\}}p(r_t=i|x_{1:t})$
\If{$t>1$ \textbf{\&} $i>\eta$}
\State Infer $\vec\lambda$ with GAS using $x_t^{(i)}-\mu_i$
\State Filter $\rho_t$
\EndIf
\State Compute $p(r_t|x_{1:t})$\Comment{The correlation of the previous step $\rho_{t-1}$ is used here}
\State Update $\mu_t,\sigma_t^2$
\EndFor
\end{algorithmic}
\end{algorithm}

Both the baseline model and its Markovian generalization assume that the parameters within each regime are constant. This might be unrealistic in many empirical cases. For example, heteroscedasticity, i.e. time-varying variance, is ubiquitous in financial time series. Since our interest here mostly focuses on the correlation of the order flow and its temporal dependencies, we consider a model where temporal correlation (i.e. $\rho$) is time-varying within the regime. This might capture the variability and temporal persistence of trading volume, which in turn depends on the available liquidity of the market.

Time-varying parameters models display typically some difficulties for estimation. Following \cite{cox}, we consider the class of {\it observation-driven models} where the parameters are unconditionally random variables, but  evolve in time based on some nonlinear deterministic function of past observations.

In particular, we consider the class of Score-Driven models introduced by \cite{Score-Driven1} and \cite{Harvey}, which assume that the dynamics of the time-varying parameter(s) is autoregressive with an innovation term depending on the so-called score\footnote{We remind that the score is the derivative of the log-likelihood with respect to the parameter(s).}. The score is then re-scaled by the inverse of the Fisher matrix\footnote{The Fisher matrix is defined as $\mathcal{I}_{t|t-1} = \E_{t|t-1}[\nabla_t^T\nabla_t]$ where $\nabla_t$ is defined in Eq. \ref{eq: II}.}, which is used to modulate the importance of the innovation according to the concavity of the log-likelihood. The intuition is simple: the scaled score adjusts the value of the parameter(s) in order to maximize the likelihood of the observed data. It is worth noticing that many standard models in financial econometrics, such as the GARCH, ACD, MEM, etc., are special cases of score-driven models (see \url{www.gasmodel.com} for more details).

We extend the MBO model by promoting the correlation coefficient $\rho$ to a time-varying parameter $\rho_t$ described by the Score-Driven version of the AR(1) process (see \cite{Score-Driven_AR}). We name such an extension as MBOC. We then introduce an online method to estimate both the time-varying parameter $\rho_t$ and the regime characteristics, namely the mean $\theta_R$ characterizing the regime and the run length $r_t$.

More specifically, within a regime $R$, the data generating process is assumed to be
\begin{align}\label{eq21}
    x_t = \rho_t(x_{t-1}-\theta_{R})+\theta_{R}+u_t,\quad u_t\sim\mathcal{N}(0,\sigma^2),
\end{align}
where $\theta_{R}$ and $\sigma^2$ are unknown. According to the Score-Driven AR(1) process, the time-varying correlation $\rho_t$ is described by the recursive relation\footnote{This specification does not guarantee that $|\rho_t|\le 1$, thus sometimes one uses a link function (e.g. an inverse logistic) which maps $[-1,1]$ in ${\mathbb R}$, see \cite{Score-Driven_AR}. In our empirical analysis, we observe that the filtered $|\rho_t|$ is larger than $1$ in less than one per thousand observations, thus we simply set a threshold $|\rho_t|\leq 1$.}
\begin{equation}
    \rho_t = \omega+\alpha s_{t-1} + \beta\rho_{t-1}
\end{equation}
where $s_t$ is the scaled score defined as 
\begin{align}\label{eq: I}
    s_t & = \mathcal{I}_{t|t-1}^{-d}\cdot \nabla_t,\quad d\in [0,1]\\
    \label{eq: II}
    \nabla_t & = \frac{\partial \log p_u(u_t)
    }{\partial \rho_t}\\
    \mathcal{I}_{t|t-1} & = \E_{t|t-1}[\nabla_t^T\nabla_t]
\end{align}
and $u_t$ is the prediction error associated with the observation $x_t$. The parameter $d$ controls the role of the Fisher matrix in modulating the role of the score in the update of the parameters. Standard choices for $d$ are $d=0,1/2,1$. Each of them lead to different dynamic scale models.
The vector of parameters $\vec\lambda = [\omega, \alpha, \beta, \sigma^2]'$ is estimated through a Maximum Likelihood Estimation method. In the analysis below, we set $d=0$, i.e. we consider a not re-scaled score. It is
\begin{align}
    s_t = \nabla_t & = \frac{u_t}{\sigma^2}(x_{t-1}-\theta_{R}).
\end{align}
Then the UPM term becomes
\begin{align}
    p(x_{t+1}|x_t^{(r_t)})
    & = \mathcal{N}\big(\mu_{r_t}+\rho_t(x_t-\mu_{r_t}),\sigma^2+\sigma_{r_t}^2\big),
\end{align}
where the posterior parameters are 
\begin{align}\label{eq: post params MBOC}
    \mu_{r_t} = \frac{b_{r_t}+\frac{\mu_0}{\sigma_0^2}}{a_{r_t}+\frac{1}{\sigma_0^2}}\quad\text{and} \quad\sigma_{r_t}^2 = \Big(a_{r_t}+\frac{1}{\sigma_0^2}\Big)^{-1}\quad\text{for}\quad r_t\in\{1,...,t\},
\end{align}
and
\begin{align}
    a_{r_t} & = \frac{1}{\sigma^2}+\frac{(r_t-1)(1-\rho_t)^2}{\sigma^2(1-\rho_t^2)}\\
     b_{r_t} & = \begin{cases}\frac{x_{t}}{\sigma^2},\quad\text{for}\quad r_t=1\\[10pt]
    \frac{x_{t-1}}{\sigma^2}+\frac{(1-\rho_t)(x_t-\rho_t x_{t-1})}{\sigma^2(1-\rho_t^2)},\quad\text{for}\quad r_t=2\\[10pt]
        \frac{x_{t+1-r_t}}{\sigma^2}+\frac{(1-\rho_t)^2\sum_{i=t+2-r_t}^{t-1}x_i+(1-\rho_t)(x_t-\rho_t x_{t+1-r_t})}{\sigma^2(1-\rho_t^2)},\quad\text{for}\quad r_t\in\{3,...,t\}
    \end{cases}.
\end{align}
The vector of parameters $\vec \lambda$ is estimated at each time-step within the time window associated with the most likely regime after we demean the data with the posterior mean see Eq. (\ref{eq: post params MBOC}). In particular at each time step $t>1$ we find 
\begin{equation*}
    i = \argmax_{i\in\{1,...,t\}}p(r_t=i|x_{1:t})
\end{equation*}
and we consider the {\it demeaned data set} $x_t^{(i)}-\mu_i = \{x_{t+1-i}-\mu_i,...,x_t-\mu_i\}$ in which we infer $\vec \lambda$ and we filter $\rho_t$ with the use of the Score-Driven model.
In order to robustify the algorithm and accomplish better Mean Squared Error (MSE) (see section \ref{sec: Online Prediction of Order Flow} for more details) we define a threshold value $\eta$, then we filter the time-varying
correlation and we infer the variance (see Algorithm \ref{alg: MBOC} for more details on the inference procedure of the MBOC model) whenever $i>\eta$. In that way, the demeaned data set contains at least $\eta$ data points. The threshold value is a hyperparameter that is tuned in a preliminary phase, see below for the implementation details. 

\section{Model Estimation and Empirical Analysis}\label{sec: Online Prediction of Order Flow}

We estimate the three models, BOCPD, MBO, and MBOC, on the time series of aggregated order flow $x_t$ of TSLA and MSFT, both for the 1 and the 3 min aggregation time scale. The application of the models requires a careful choice of several hyperparameters. For both TSLA and MSFT, the prior value of the mean for all models is set to $\mu_0 = 0$ shares due to the symmetry between buy and sell orders. The tuning of the other hyperparameters is obtained by minimizing the MSE in the first day of each month and each stock. For TSLA, for all the models, the prior value of the variance of the mean is set to $\sigma_0^2 =  10^7$  shares$^2$, while for MSFT $\sigma_0^2 =  15\cdot 10^7$ shares$^2$. The hazard rate is set to $h = \frac{1}{80}$. For  both BOCPD and MBO the known variance is set to $\sigma^2 = 10^8$ shares$^2$ for TSLA while for MSFT $\sigma^2 =  15\cdot 10^8$ shares$^2$. The same values for TSLA and MSFT are being used for the initial variance $\sigma_i^2$ of the MBOC model (see Algorithm \ref{alg: MBOC}). Moreover, the initial correlation $\rho_1$ is set to 0.2 for TSLA and 0.3 for MSFT while the initial parameters of the Score-Driven dynamics are set to $\vec\lambda = [0.08,0.02,0.05,10^8]'$ and the $\eta$ value is set equal to 20 shares for the 1 minute and 10 shares for the 3 minute data set. The hyperparameter $\eta$ is being tuned in such a way in order to obtain the best MSE. Finally, for the constant correlation coefficient $\rho$ of MBO we have tested different specifications. In the table below we will report the results for three of them, showing that the predictive capacity of the model slightly depend on it.

\subsection{Model comparison}
We present here the results of an online prediction study for order flow data by using BOCPD, MBO, and MBOC models introduced in Section \ref{sec: Bayesian Online Change-Point Detection Algorithms}, and we compare their performances by computing the Mean Squared Error for the predictive mean of each model. The three models are then compared with the ARMA(1,1) model, estimated on the whole time period by assuming the absence of regimes. As such, the ARMA(1,1) model represents a natural benchmark to test whether including regime-switching dynamics does improve the forecasting of order flow.

The predictive mean $\hat{\mu}_t$ at time $t$ as one step ahead forecast is defined by using observations up to time $t$ and to predict out-of-sample the realization at time $t+1$. That is
\begin{align}
    \hat{\mu}_{t} & = \sum_{r_t} p(x_{t+1}|x_{1:t},r_t,)p(r_t|x_{1:t})\nonumber \\
    & = \sum_{r_t} \mu_{r_t} p(r_t|x_{1:t})
\end{align}
where $\mu_{r_t}$ is defined in Equations (\ref{eq: post params BOCPD}), (\ref{eq: post params MBO}), and (\ref{eq: post params MBOC}) for BOCPD, MBO and MBOC model respectively. 
Then, the MSE can be computed as
\begin{equation}
    \text{MSE} = \frac{1}{T}\sum_{t=1}^T(\hat{\mu}_{t-1}-x_t)^2.
\end{equation}

Table \ref{tab1} shows the MSE of the three aforementioned models along with the ARMA(1,1) for both TSLA and MSFT stocks. As mentioned above, we consider three different values of the correlation coefficient $\rho$ in the MBO model. We observe that the MBOC model outperforms all competitors.
In particular, the proposed models (MBO and MBOC) outperform systematically the ARMA(1,1) benchmark, while the baseline BOCPD model displays comparable performances. Finally, notice that the role of the hyperparameter $\rho$ for the MBO model is relatively marginal.

In conclusion, the online prediction study with order flow data suggests that regime-switching models accounting for a Markovian correlation structure outperform both the baseline BOCPD model and the benchmark. The MBOC model displays the best forecasting performance and high flexibility in data description. In the following Sections, we exploit such flexibility in modeling regime-switching dynamics in the presence of time-varying correlations to empirically show a clear connection between regimes for aggregated order flows and the market impact of associated trades (likely including metaorders).

\begin{center}
\begin{tabular}{c||c|c|c||c|c|c}
\hline
 & \multicolumn{3}{c||}{TSLA} & \multicolumn{3}{c}{MSFT} \\
\cline{1-7}
  $\rho$ & $0.1$ & $0.2$ & $0.3$ & $0.2$ & $0.3$ & $0.4$\\
\hline\hline
ARMA & 0.908 & 0.908 & 0.908 & 0.860 & 0.860 & 0.860\\

BOCPD & 0.907 & 0.907 & 0.907 & 0.878 & 0.878 & 0.878\\

MBO & 0.895 & 0.896 & 0.911 & 0.835 & 0.834 & 0.844\\
  
   \textbf{MBOC} & \textbf{0.890} & \textbf{0.890} & \textbf{0.890} & \textbf{0.831} & \textbf{0.831} & \textbf{0.831} \\  
  \hline
\end{tabular}
\captionof{table}{Comparison of out of sample one-step-ahead MSE of ARMA(1,1), BOCPD, MBO and MBOC. The correlation $\rho$ is the one used in MBO.
Data refer to TSLA and MSFT at the 3 minute resolution.}
\label{tab1}
\end{center}

\subsection{Empirical Analysis of Identified Regimes}

\begin{figure}[t]
\centering
\includegraphics[width=1\columnwidth]{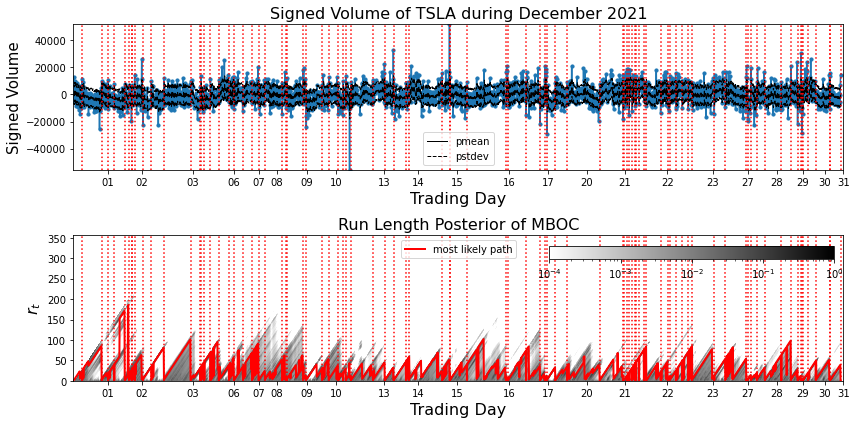}
\caption{\textbf{Top:} Order flow of Tesla at the 3 minutes aggregation time scale during December 2021. The black solid line is the predictive mean and the dashed line, the predictive standard deviation. The red dashed lines indicate the CPs the MBOC finds and as a result the regimes. The tick labels on the x-axis indicate the end of each trading day. \textbf{Bottom:}  Run length posterior of the MBOC model. The darkest the color, the highest the probability of the run length. The red line highlights the most likely path, i.e. value of $r_t$ with the largest run length posterior $p(r_t|x_{1:t})$ for each $t$.}
\label{fig2}
\end{figure}

\begin{figure}[t]
\centering
\includegraphics[width=1\columnwidth]{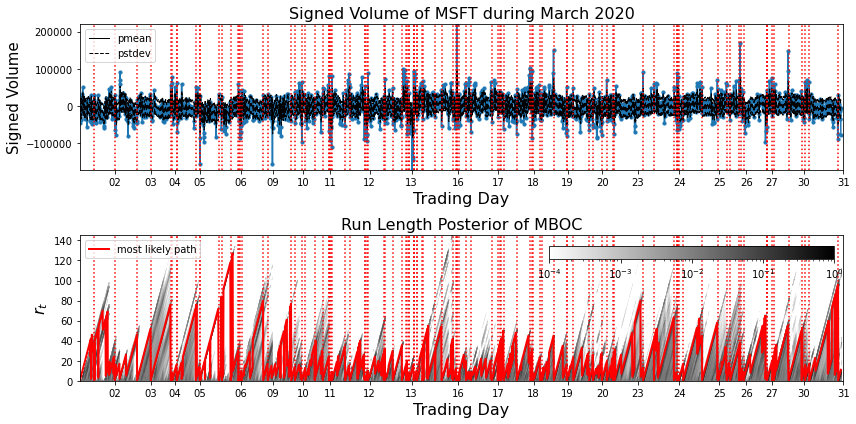}
\caption{\textbf{Top:} Order flow of Microsoft at the 3 minutes aggregation time scale during March 2020. The black solid line is the predictive mean and the dashed line, the predictive standard deviation. The vertical red dashed lines indicate the CPs the MBOC finds and as a result the regimes. The tick labels on the x-axis indicate the end of each trading day. \textbf{Bottom:}  Run length posterior of the MBOC model. The darkest the color, the highest the probability of the run length. The red line highlights the most likely path, i.e. value of $r_t$ with the largest run length posterior $p(r_t|x_{1:t})$ for each $t$.}
\label{fig3}
\end{figure}
Here we investigate the statistical properties of the identified regimes for the aggregated order flows. We consider the MBOC model because of the best performances. Let us first introduce the adopted definition of an {\it identified} regime.
\begin{definition}\label{def:regime}
Let $x_{1:T}$ be a time series and $t,s\in\N\cap[1,T]$ times with $t<s$ such that 
\begin{equation}
    \argmax_{i\in\{0,1,...,t\}}p(r_t=i|x_{1:t}) = \argmax_{i\in\{0,1,...,s\}}p(r_s=i|x_{1:s}) = 0
\end{equation}
 and for any $u\in\N\cap(t,s)$, $\argmax_{i\in\{0,1,...,u\}}p(r_u=i|x_{1:u}) \neq 0$. Then the subset $x_{t:s-1}$ of the time series $x_{1:T}$ is defined as a regime.
\end{definition}

 The top panel of Figure \ref{fig2} (\ref{fig3}) shows $x_t$ for TSLA (MSFT) aggregated every 730 (1200) executions corresponding to an average time interval of 3 minutes. The vertical red dashed lines indicate the CPs identified by MBOC, according to the definition above. Interestingly, many CPs are observed at the end of a trading day  for both stocks. On one side this is expected since overnight is a natural separation between regimes, but on the other side, this is an indication that the proposed method is able to identify regime changes.

The bottom panels of Figure \ref{fig2} and \ref{fig3} show the run length posterior of the MBOC model for the two assets. For each time (on the abscissa) the vertical axis displays in grayscale the probability that the run length has a given value (on the ordinate). Darker grey regions correspond to higher probabilities.
The red line highlights the most likely path, i.e. the value of $r_t$ with the largest run length posterior $p(r_t|x_{1:t})$ for each $t$. Finally, we also show the (one step ahead) predictive standard deviation defined as
\begin{equation}
    \hat{\sigma}_{t} = \sqrt{\sum_{r_t} \sigma_{r_t}^2 p(r_t|x_{1:t})},
\end{equation}
where $\sigma^2_{r_t}$ is as in Equations (\ref{eq: post params BOCPD}), (\ref{eq: post params MBO}), and (\ref{eq: post params MBOC}) for BOCPD, MBO, and MBOC models, respectively. 

\begin{figure}
    \centering
    \subfigure[]{\includegraphics[width=0.24\textwidth]{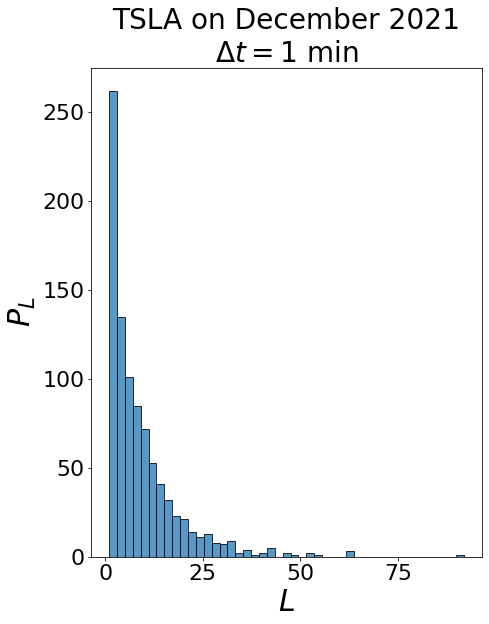}} 
    \subfigure[]{\includegraphics[width=0.24\textwidth]{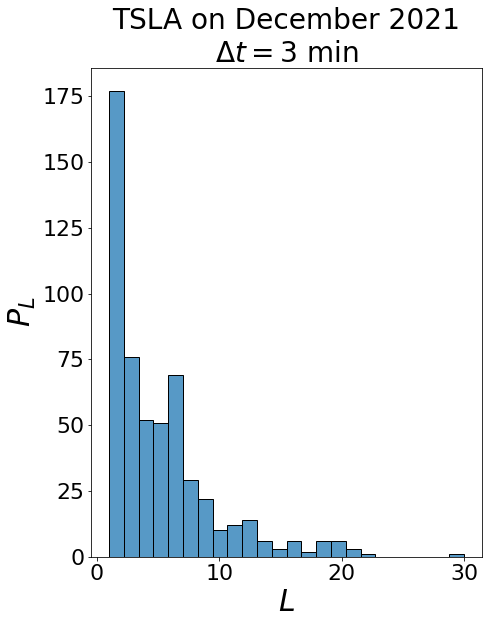}}
    \subfigure[]{\includegraphics[width=0.24\textwidth]{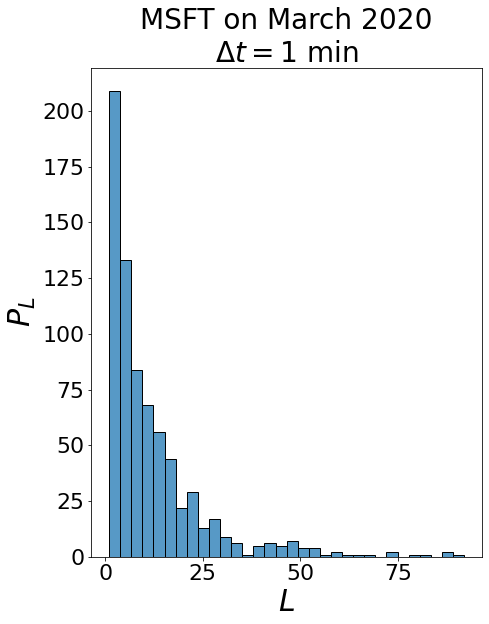}} 
    \subfigure[]{\includegraphics[width=0.24\textwidth]{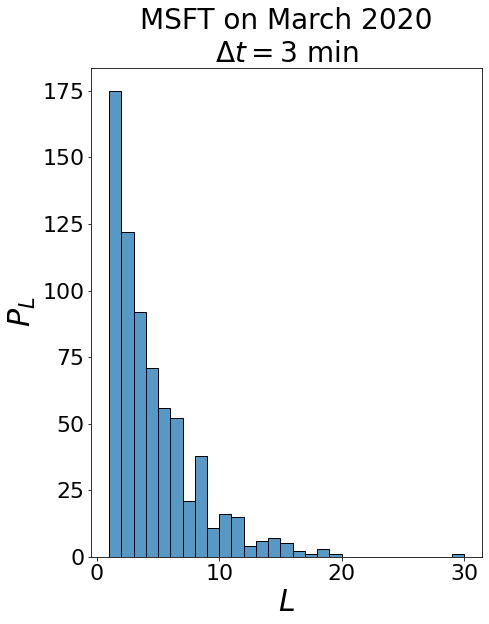}}
    \caption{Histograms of regimes length for Tesla (a) with $\Delta t=1$ min, (b) with $\Delta t=3$ min and for Microsoft (c) with $\Delta t=1$ min, (d) with $\Delta t=3$ min.}
    \label{fig4}
\end{figure}

{\bf Regime length distribution.} 
For the 1 (3) minute(s) data set, we find 911 (546) regimes for TSLA and 1394 (690) regimes for MSFT. Figure \ref{fig4} shows the histograms of the length of the detected regimes. Consistently with the constant hazard function, we find that the regime length is approximately exponentially distributed with a mean regime length of 10 for the 1 minute data set and 5 for the 3 minutes data set intervals for TSLA and 7 for the 1 minute data set and $4$ intervals for the 3 minutes data set of MSFT corresponding to 10 and 15 trading minutes for TSLA and to 7 and 12 trading minutes for MSFT respectively. The length of the regimes ranges in the interval $[1,92]$ for 1 minute and in $[1,30]$ for 3 minutes for TSLA  while for MSFT in $[1,65]$ for 1 minute and in $[1,30]$ for 3 minutes for MSFT.

{\bf Gaussianity inside regimes.} The main assumption of both BOCPD and MBO is that the variable $x_t$ is Gaussian within each regime, with constant parameters. For MBOC, we expect that $x_t$ is only conditionally Gaussian, because of the time-varying autocorrelation, but not unconditionally over the whole period and only approximately within a regime. The Jarque-Bera (JB) test rejects
the Gaussianity hypothesis of unconditional aggregated order flow $\{x_t\}_{t=1,\ldots,T}$ at a $1\%$ significance level, for 
both stocks and time scales.
We then perform the JB test within each regime detected by the MBOC model. For the 3 minute time scale, we cannot reject the null hypothesis at $5\%$ confidence level for {\bf $94\%$ ($95\%$)}  of the regimes for TSLA (MSFT). When we consider the 1 minute time scale, the frequency of rejection is {\bf $86\%$ ($87\%$)}. These findings support the choice of MBOC to identify regimes with order flows as approximately Gaussian within each regime.

{\bf Autocorrelation of residuals inside regimes.} As a final model checking we test for the lack of serial correlation in the residuals of our model within each regime. We have seen above that, coherently with the literature, $v_t$ is strongly autocorrelated. Following \cite{ar:2005}, our assumption is that this correlation is driven by the presence of regimes, which in turn are likely associated with metaorders. We thus apply the Ljung-Box test to the residuals in each regime. For the 3 minute time scale, we cannot reject the null hypothesis of uncorrelated residuals for {\bf $98\%$ ($99\%$)} of the regimes of TSLA (MSFT), with $5\%$ confidence level. For the 1 minute time scale, the frequency of rejection is {\bf $97\%$ ($93\%$)}.
It is possible to conclude that the MBOC model captures most of the serial correlation of aggregated order flow. Notice that, according to the model, the unconditional slow decay of the autocorrelation of order flow observed in the literature (see also Fig. \ref{fig1}) is due largely to regime-switching dynamics and, only partially, to Markovian temporal dependencies.

\section{Price impact during order flow regimes}\label{sec:impactduring}

In this Section we empirically study the average price dynamics inside a detected order flow regime and we measure the relation between the total price change and the net volume exchanged in the same regime.

\subsection{Price as a function of time  inside an order flow regime}

 As said, we first study the average price dynamics during an order flow regime. This type of analysis mirrors the one performed, for example, by \cite{Bacry2015,r:2015} which studied the average price dynamics during the execution of a metaorder. Using labeled data allowing to identify when an institutional investor executed a metaorder, these papers find that (i) the average price dynamics is correlated with the conditioning metaorder sign, i.e. the price increases (decreases) when a buy (sell) metaorder is executed; (ii) the price dynamics is concave in time, i.e. the price increases faster at the beginning of a buy metaorder and slowly toward the end. 

Here we take a step forward by asking what is the average price dynamics during a regime of aggregated order flow detected with the MBOC model. To this end, for each detected regime $R$ (see Definition \ref{def:regime}), characterized by an initial time $t_R$ and a final time $s_R>t_R$, we denote with 
$$\epsilon_R =\sign\left(\sum_{t_R\le t <s_R}x_t\right)
$$ 
the sign of the order flow in the regime, being equal to $+1$ ($-1$) when the regime is dominated by the volume of buyer (seller) initiated trades. Since during a metaorder execution we expect a significant net imbalance of buy or sell volume, we will consider subsets of regimes for which
\begin{equation}\label{eq:Theta condition}
   Z_R:=\left| \frac{\sum_{t_R\leq t< s_R}x_t}{\sum_{t_R\leq t< s_R}|x_t|}\right |>\Theta,~~~~~~~~~~0\le \Theta <1
\end{equation}
i.e. when the difference between buy and sell volume divided by their sum is larger than $\Theta$. Notice that when $\Theta=0$ the subset coincides with the entire set of regimes identified by the MBOC. Appendix \ref{appA1} reports the number of regimes in the different subsets.

Indicating with $p_t$ the log-price of the last transaction in the interval labeled by time $t$, 
we compute the log-price change between the beginning of the regime  and $t_R+k$, where $0\le k< s_R-t_R$ and we take the average 

\begin{equation}\label{eq:impact}
    \mathcal{I}^{\Theta}(k) = \E_R\left[\epsilon_R(p_{t_R+k}-p_{t_R-1})|t_R+k< s_R,Z_R>\Theta\right].
\end{equation}
With $\E_R[\cdot]$, we denote the sample average over the regimes, i.e. that $t_R$ is the first interval of a regime,  and the conditioning restricts it to those regimes for which the observation at $t_R+k$ is in the same regime as the one at $t_R$ as well as to those regimes that satisfy condition in Eq. (\ref{eq:Theta condition}).

\begin{figure}
    \centering
     \subfigure[]{\includegraphics[width=0.24\textwidth]{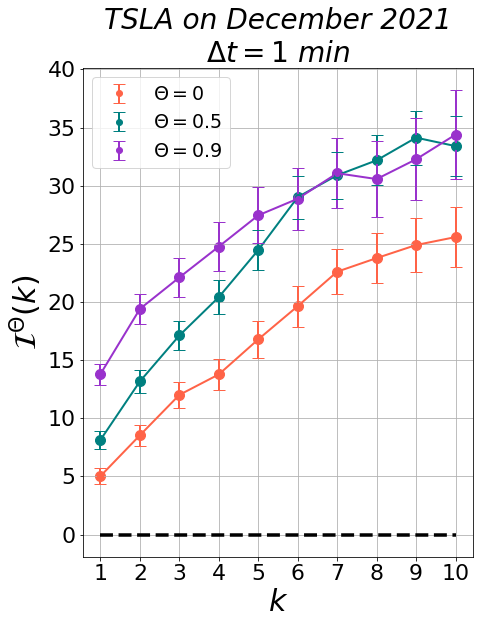}}
    \subfigure[]{\includegraphics[width=0.24\textwidth]{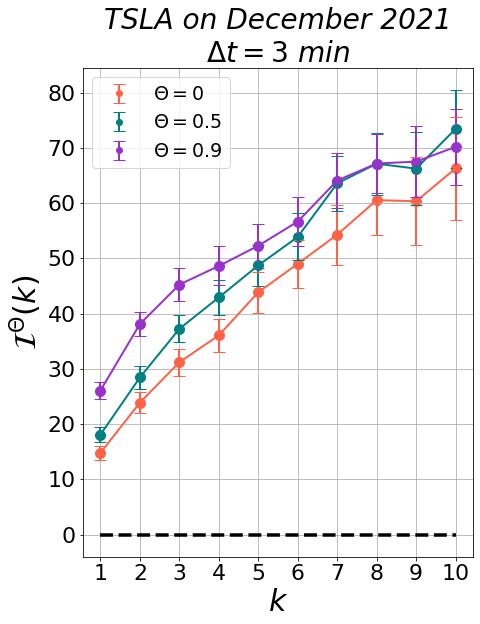}}
    \subfigure[]{\includegraphics[width=0.24\textwidth]{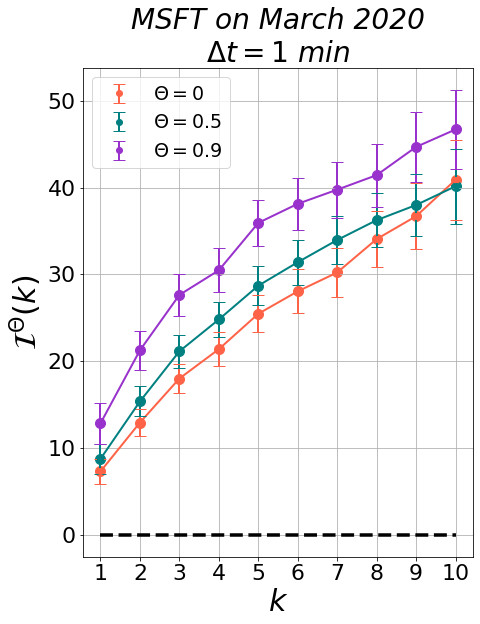}}
    \subfigure[]{\includegraphics[width=0.24\textwidth]{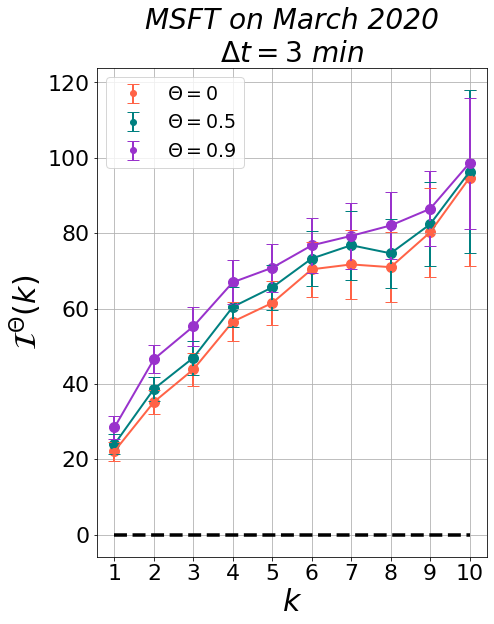}}    
    \caption{Market Impact $\mathcal{I}^{\Theta}(k)$ defined in Eq. (\ref{eq:impact})  for Tesla (a) with $\Delta t=1$ min, (b) with $\Delta t=3$ min and for Microsoft (c) with $\Delta t=1$ min, (d) with $\Delta t=3$ min. The plotted quantities are in basis points. Error bars are standard errors. The black dashed line indicates the $y=0$ axis.}
    \label{figimp}
\end{figure}

Figure \ref{figimp} shows the impact function $\mathcal{I}^{\Theta}(k)$ as a function of $k$ for the two stocks and the two time scales when $\Theta=0, 0.5$, and $0.9$. Error bars are standard errors in each bin. We notice that in all cases impact is positive and increasing. This is somewhat expected since we are implicitly conditioning on the sign of order flow in the {\it whole} regime, thus the observed behavior is coherent with the known correlation between aggregated order flow and contemporaneous price change, see \cite{em:09,PhysRevE.97.012304}. Interestingly the price dynamics is a concave function of time, similarly to what is observed when conditioning on metaorders execution instead of on order flow regimes. Moreover the degree of concavity increases with $\Theta$. 
Clearly the concavity is not expected by the mere fact that the regime is characterized by a net order flow sign, while it could instead be explained by the Transient Impact Model of \cite{Bouchaud04} or by the LLOB model of \cite{LLOB}, which predicts a concave average price temporal profile when the order flow has a non-zero average as during a metaorder execution.

\subsection{Price impact as a function of volume} 

Finally, we study the relation between the total price change in a regime and the total net volume in the same time span. A large body of empirical literature \cite{BARRA,Almgren2005DirectEO,SRILforOptions,PhysRevX.1.021006,r:09,r:2015,ar:2013} have shown that on average the total price impact during a metaorder execution scales with a sublinear power law of metaorder volume, a relation well fit by a power law with an exponent ranging in $[0.4,0.7]$. This is the celebrated {\it square root impact law}. It is therefore natural to investigate empirically the relation between the same two quantities within a regime identified with our method.

To this end, defining $\Delta p_R = p_{s_R} - p_{t_R}$ we consider the non-linear regression:  
\begin{equation}\label{eq: regression}
    \epsilon_R\Delta p_R = A(\epsilon_R\sum_{t_R\leq t<s_R}x_t)^{\gamma}+noise.
\end{equation}

Since the measurement of market impact is notoriously very noisy, we have performed the estimation both on the original dataset and on a dataset where potential outliers are removed. In the latter approach, we used the standard procedure of considering outliers datapoints corresponding to regimes whose price change is smaller (larger) than the first (third) quartile minus (plus) 1.5 times the interquartile range (see Appendix \ref{AppA2} for details).  

Table \ref{tab2} reports the estimated parameters when outliers are removed, while Appendix \ref{AppA2} reports the results for the entire dataset and presents the scatter plots of the data and the fitted curve. Both Tables indicate that the exponent $\gamma$ is smaller than one and for the data without outliers it is remarkably close to $0.5$, as postulated by the square root law.

\begin{center}
\begin{tabular}{c||c c c c|c c c c}
\hline
& \multicolumn{8}{c}{TSLA}\\
\cline{2-9}
 & \multicolumn{4}{c|}{$\Delta t=1$ min} & \multicolumn{4}{c}{$\Delta t=3$ min}\\
$\Theta$ & $A$ & SE of $A$ & $\gamma$ & SE of $\gamma$ & $A$ & SE of $A$ & $\gamma$ & SE of $\gamma$ \\
\hline\hline
0 & $0.121$ & $0.05$ & $0.592$ & $0.041$ & $0.298$ & $0.139$ & $0.52$ & $0.045$\\

0.5 & $0.159$ & $0.066$ & $0.567$ & $0.041$ & $0.284$ & $0.133$ & $0.528$ & $0.045$\\

0.9 & $0.172$ & $0.084$ & $0.552$ & $0.049$ & $0.387$ & $0.179$ & $0.504$ & $0.044$\\  
  \hline
  \hline
& \multicolumn{8}{c}{MSFT}\\
\cline{2-9}
 & \multicolumn{4}{c|}{$\Delta t=1$ min} & \multicolumn{4}{c}{$\Delta t=3$ min}\\
$\Theta$ & $A$ & SE of $A$ & $\gamma$ & SE of $\gamma$ & $A$ & SE of $A$ & $\gamma$ & SE of $\gamma$ \\
\hline\hline
0 & $0.498$ & $0.209$ & $0.4$ & $0.038$ & $0.365$ & $0.203$ & $0.458$ & $0.048$\\

0.5 & $0.469$ & $0.209$ & $0.408$ & $0.04$ & $0.314$ & $0.185$ & $0.47$ & $0.05$\\

0.9 & $0.334$ & $0.144$ & $0.444$ & $0.038$ & $0.332$ & $0.199$ & $0.47$ & $0.051$\\  
  \hline
\end{tabular}
\captionof{table}{Estimated parameters and their standard errors (SE) of the regression of Eq. (\ref{eq: regression}). Results refer to data with outliers removed.
}
\label{tab2}
\end{center}

Clearly these results are preliminary and should be validated on larger panels of stocks, also pooling them together with the usual rescaling by daily volatility and volume. However we find these results very encouraging and suggestive of a relation between the identified regimes and the execution of metaorders.

\section{Online prediction of order flow and market impact} 

The possibility of performing an online detection of regimes and regime changes in the order flow opens the question of how to use this information to predict subsequent order flow and price changes.  In the PPM regimes are independent, hence in forecasting future values only data from the current regime are useful, while older data add noise to the prediction. This idea will be used to build online predictions of order flow. Additionally, through market impact, price dynamics is correlated with order flow. Thus a proper modeling of order flow is useful to forecast future price.

Since we have seen in the last Section that  order flow sign of a regime correlates with contemporaneous price change, we can ask the question of whether the knowledge that a new regime in order flow has just started allows to predict the future order flow and, more importantly, the future price dynamics.  
 Consistently with the results of Section \ref{sec: Online Prediction of Order Flow} 
 showing that the MBOC model outperforms the competitors (BOCPD, MBO, ARMA) in one step ahead prediction of order flow, here we focus our analyses on the regimes identified by MBOC. It is important however to stress that qualitatively similar results are obtained with the other two simpler models, BOCPD and MBO. In other words the relation between price dynamics and aggregated order flow is importantly understood by using regimes, while the choice of the specific regime shift model improves the short-term prediction of aggregated order flow. 
 However, as shown in in the appendix \ref{appB}, the MBOC method achieves higher predictability wrt the other CP detection methods.

To better understand the role of regimes in prediction, let consider the following argument. If the data generating process of order flow is truly consistent with a product partition model (i.e. independent regimes), the knowledge of the data of the previous regimes is not useful for prediction. Thus in this case, it is better to use only the data in the current regime and its learned statistical properties. However, even if we are relatively sure that a new regime has just started, its parameters could be quite uncertain at the beginning. Thus, in order to form a forecast, it is better to wait for few observations into a new regime. Moreover, if many regimes are very short (e.g. composed by one or two intervals), as observed empirically in Figure \ref{fig4}, it might be better to build predictions after the observation of a few intervals in a regime.

{\bf Online order flow prediction.} Following the above argument, we adopt the following procedure. Whenever we detect a CP in an online fashion for the time series of order flow, we measure the correlation

\begin{equation}
    \mathcal{I}^{(1)}_\epsilon(k) = \E_R[\sign(x_{t_R}) \cdot \sign(x_{t_R+k})]~~~~~~k=1,2,...
\end{equation}
where $\E_R$ indicates that we are conditioning on the fact that $t_R$ is the first observation of a new regime.
Notice that (i) the correlation is extended to values of $k$ possibly beyond the end of the detected regime (at time $s_R$); (ii) we do not consider the case $k=0$ since in this case the correlation is trivially equal to $1$. The superscript $^{(1)}$ in the above expression means that we take the sign of the aggregated order flow in the first interval (see below for an extension).
\begin{figure}[!htb]
    \centering
    \subfigure[]{\includegraphics[width=0.24\textwidth]{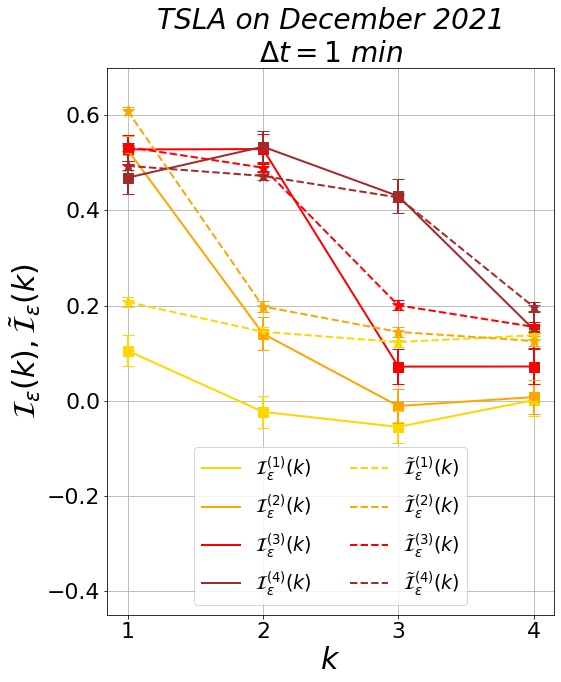}}
    \subfigure[]{\includegraphics[width=0.24\textwidth]{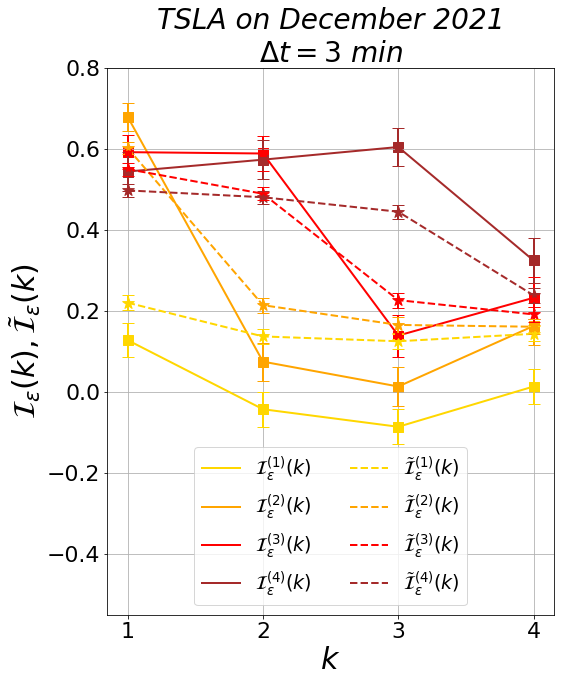}}  
    \subfigure[]{\includegraphics[width=0.24\textwidth]{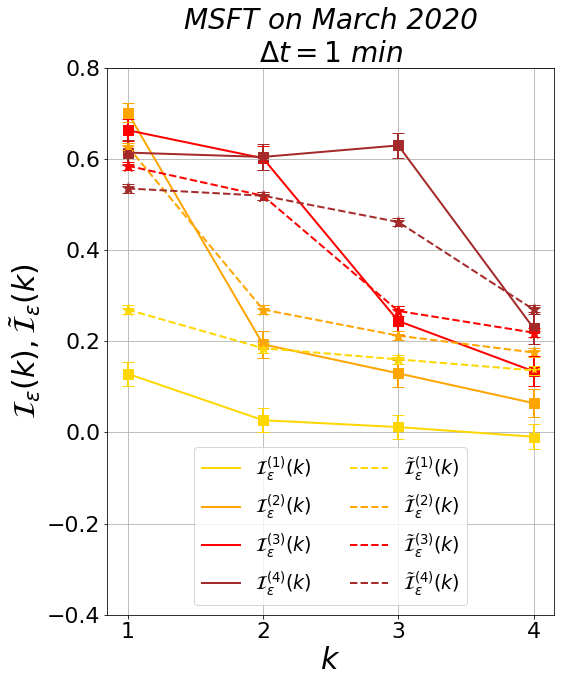}}
    \subfigure[]{\includegraphics[width=0.24\textwidth]{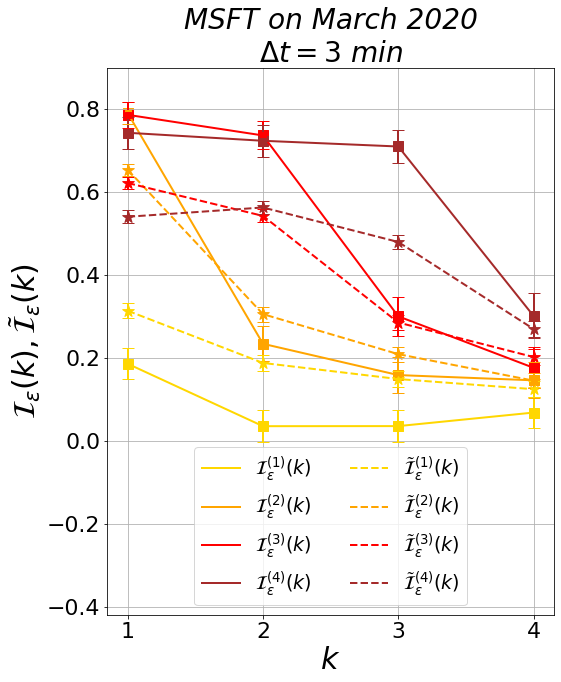}}  
    \caption{Online order flow prediction for Tesla (a) with $\Delta t = 1$ min, (b) with $\Delta t = 3$ min and for Microsoft (c) with $\Delta t = 1$ min, (d) with $\Delta t = 3$ min.  The plotted quantities are defined in Eqs. (\ref{eq:Iepsm}) and (\ref{eq:Iepsmu}) and error bars are standard errors.}
    \label{fig:onlineOF}
\end{figure}
The continuous yellow line in  Figure \ref{fig:onlineOF} shows that $\mathcal{I}^{(1)}_\epsilon(k)$ is a poor predictor of order flow. To better quantify this statement, the dashed yellow line in the figure is the unconditional correlation 
\begin{equation}\label{eq:Iepsu}
    \mathcal{\tilde I}^{(1)}_\epsilon(k) = \E[\sign(x_{t}) \cdot \sign(x_{t+k})]~~~~~~k=1,2,...
\end{equation}
which makes no use of regime detection (and for this reason the expectation does not have the subscript $_R$; the tilde refers to the expectation without considering regimes). Clearly, the unconditional correlation is larger than the conditional one.

As said above, one of the reasons of the comparable performance of $\mathcal{I}^{(1)}_\epsilon$ with respect to $\mathcal{\tilde I}^{(1)}_\epsilon$ is the fact that there are many regimes of length one and also that the sign of the new regime, $\epsilon_R$, might be poorly measured by the sign of the first interval $\sign(x_{t_R})$. A better option is to wait few more intervals within the regime before building the predictor. Thus, defining $m=1,2,..$ the number of intervals in a regime we wait before forming the prediction, we introduce the correlation
\begin{equation}\label{eq:Iepsm}
    \mathcal{I}^{(m)}_\epsilon(k) = \E_R\left[\sign\left(\sum_{t=t_R}^{t_R+m-1}x_{t}\right) \cdot \sign(x_{t_R+m-1+k})\right]~~~~~~k=1,2,...
\end{equation}
which is the correlation between the sign of the order flow in the first $m$ intervals of a regime and the sign of the order flow in an interval $k$ steps after these $m$ intervals. Notice that we are not conditioning on the fact that $t_R+m-1+k$ is in the same regime as $t_R$, so the two observation could belong to different regimes. However, we condition on the fact that $t_R+m-1$ is in the same regime as $t_R$.  Similarly we use as a benchmark case the predictor
\begin{equation}\label{eq:Iepsmu}
    \mathcal{\tilde I}_{\epsilon}^{(m)}(k) = \E\left[\sign\left(\sum_{s=t}^{t+m-1}x_s\right)\cdot \sign(x_{t+m-1+k})\right]~~~~~~k=1,2,...\:.
\end{equation}
The orange, red, and dark red continuous lines in Figure \ref{fig:onlineOF} show $\mathcal{I}^{(m)}_\epsilon(k)$ for $m=2,3,4$ respectively, while the corresponding dashed lines refer to $\mathcal{\tilde I}^{(m)}_\epsilon(k)$ . We observe that the correlations based on regimes are larger than the corresponding ones without regimes, especially for large $m$. This empirical evidence indicates that the knowledge of the order flow regimes improves the short-term predictability of order flow.

{\bf Online market impact prediction.} We now consider the prediction of price change based on the knowledge of being in a regime of order flow. To this end we introduce the online impact
\begin{equation}\label{eq:Idpm}
    \mathcal{I}^{(m)}_{\Delta p}(k) = \E_R\left[\sign\left(\sum_{t=t_R}^{t_R+m-1}x_{t}\right) \cdot (p_{t_R+m-1+k}-p_{t_R+m-1})\right]~~~~~~k=1,2,...
\end{equation}
which is the correlation between the sign of the total order flow in the $m$ initial intervals of a regime and the subsequent price change over $k$ intervals. Compared to Eq. (\ref{eq:impact}), two important differences are worth to be highlighted. First, the sign inside the expectation is taken only on the aggregated order flow of the $m$ intervals used to build the prediction, while in Eq. (\ref{eq:impact}) $\epsilon_R$ considers the sign of the whole regime and therefore is non-causal. Second, in $\mathcal{I}^{(m)}_{\Delta p}(k)$ we do not condition on $s_R-t_R>k$ as in Eq. (\ref{eq:impact}) since after having observed $m$ intervals in a regime we do not know when the regime is going to end. In other words, for a given $k$, we take the average both on cases when $t_R$ and $t_R+k$ belong to the same regime and when they do not. Finally, as before we use as a benchmark an impact predictor that is based on the sign of the order flow,
\begin{equation}\label{eq:Idpu}
    \mathcal{\tilde I}^{(m)}_{\Delta p}(k) = \E\left[\sign\left(\sum_{s=t}^{t+m-1}x_s\right)\cdot (p_{t+m-1+k}-p_{t+m-1})\right]~~~~~~k=1,2,...\:.
\end{equation}
When $m=1$ the quantity $\mathcal{\tilde I}^{(m)}_{\Delta p}(k)$ becomes the response function widely investigated in the market impact literature, mainly in transaction time, see \cite{Bouchaud04,em:09}. We choose this more general definition in order to make a fairer comparison between impact predictors using the same number of past order flow observations.

\begin{figure}[!htb]
    \centering
    \subfigure[]{\includegraphics[width=0.24\textwidth]{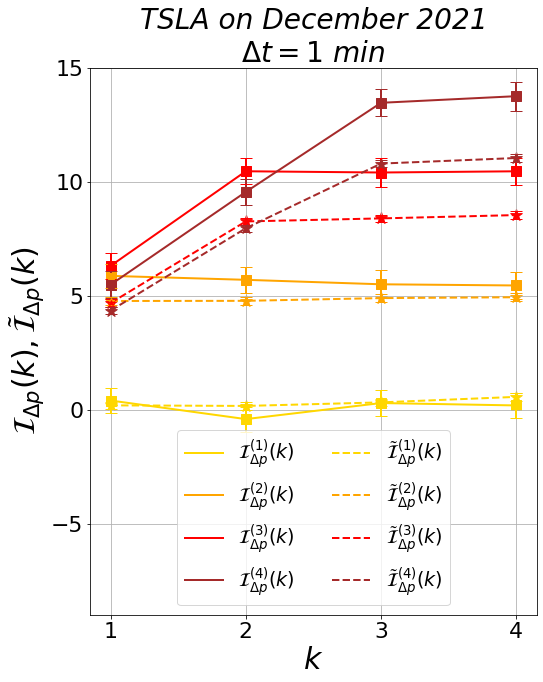}}
    \subfigure[]{\includegraphics[width=0.24\textwidth]{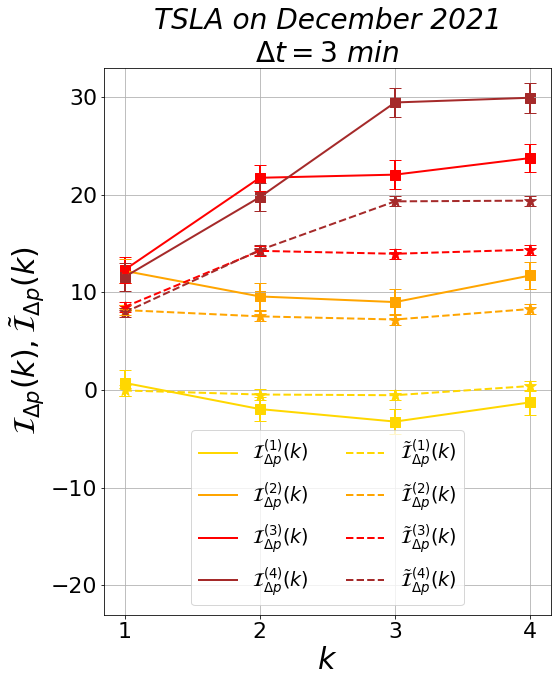}}  
    \subfigure[]{\includegraphics[width=0.24\textwidth]{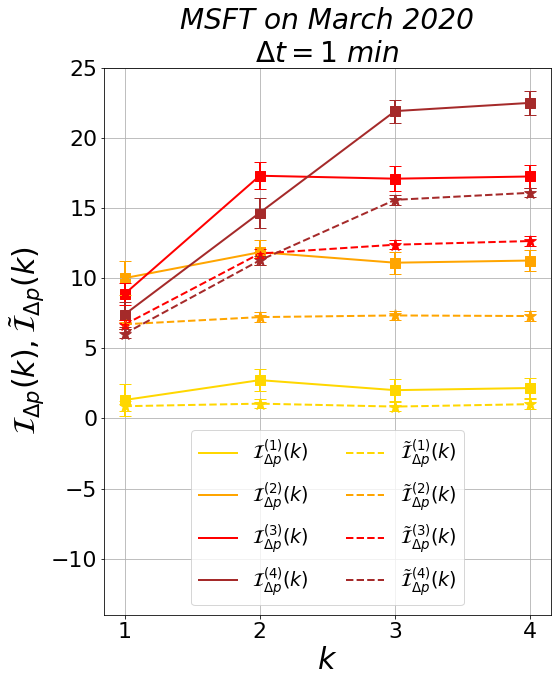}}
    \subfigure[]{\includegraphics[width=0.24\textwidth]{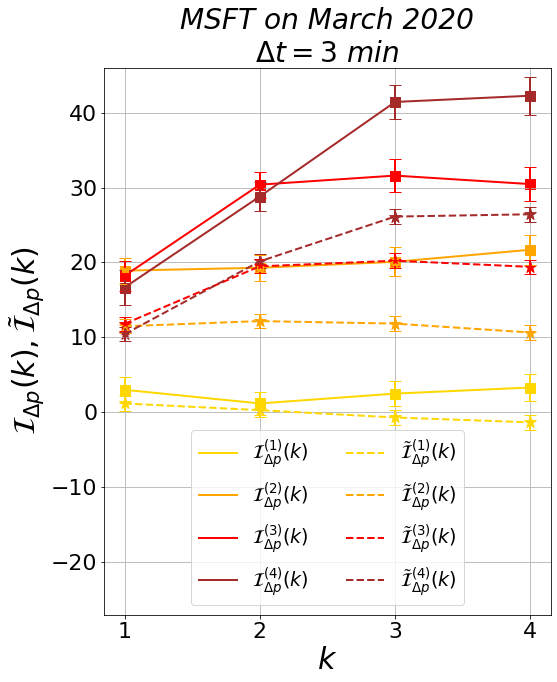}}  
    \caption{Online market impact prediction for Tesla (a) with $\Delta t = 1$ min, (b) with $\Delta t = 3$ min and for Microsoft (c) with $\Delta t = 1$ min, (d) with $\Delta t = 3$ min.  The plotted quantities (in basis points) are defined in Eqs. (\ref{eq:Idpm}) and (\ref{eq:Idpu})  and error bars are standard errors.}
    \label{fig:onlineMI}
\end{figure}

Figure \ref{fig:onlineMI} shows these different quantities for online market impact prediction, considering both stocks and both time intervals. It is evident that as soon as $m>1$,  $\mathcal{I}^{(m)}_{\Delta p}(k)$ (orange to dark red lines) is much larger than the corresponding response function $\mathcal{\tilde I}^{(m)}_{\Delta p}(k)$ (dashed orange to dark red lines) which does not make use of regimes. Moreover, the larger $m$, the larger the correlation between the order flow sign and the future price change, in all four investigated cases.  
Thus the (online) knowledge that a regime has started provides a significant additional forecasting power to future price change with respect to the response function, which is an unconditional cross-correlation between current order flow and future price change.

\section{Conclusion}
In this work, we proposed the use of Bayesian Online Change Point Detection Methods to identify (in a real-time setting) regimes in time series of aggregated order flow of financial assets. Since the existing methods make very strong assumptions on the data generating process, in particular for what concerns the serial correlation of data within each regime, we proposed here two extensions of the regime detection algorithm: the first one assumes a Markovian dynamics inside each regime, while the second one makes use of an observation driven dynamics based on the score-driven mechanism. As shown by the recent econometric literature, the score driven approach is extremely flexible also as a filter of a misspecified dynamics (tantamount to GARCH). The companion paper (\cite{tsaknaki}) provides more methodological details of this new class of models by discussing different specifications where other parameters (e.g. the variance) are time-varying within each regime. The analysis of two liquid stocks traded in the NASDAQ market shows that the new algorithms presented here, particularly the latter, outperform the baseline model in out-of-sample forecasting. In general, we find that the regime-switching methods outperform standard econometric time series models like ARMA(1,1). Moreover, a careful model checking shows that the algorithm outputs well specified regimes both in terms of Gaussianity of data and of lack of serial correlation of residuals, within each regime.

From the financial point of view, the identification of weakly autocorrelated regimes in the order flow time series suggests that the observed unconditional long memory might be explained by regime switching. This is in line with the  mechanism proposed by \cite{ar:2005} who connected the long memory to order splitting by heterogeneous institutional investors. It is natural at this point to try to identify the detected regimes with time periods when one or a few institutional investors are trading a large order. Of course, we do not have any empirical evidence in support of this idea which, at this point, can be considered as a conjecture to be tested with suitable data (for example those used by \cite{r:2015} or \cite{c:23}).

The paper shows how the online identification of regimes can be used to significantly improve the forecasting of order flow and of price dynamics. Using the knowledge of the order flow during the current regime provides better predictions when compared with methods using unconditionally the past history of order flow. We foresee that such improvement could be fruitfully used in several financial applications, such as optimal trading, market making, and alpha signal detection. Similarly, if our interpretation above is correct, the online regime detection method could be used to statistically identify the execution of a large institutional execution from anonymous market data.

\section*{Declarations}

\subsection*{Acknowledgement}
This paper is funded by European Union Next Generation EU with the grant PNRR IR0000013 ‘‘SoBigData.it".

\subsection*{Competing interests}
The authors report there are no competing interests to declare

\subsection*{Data availability}
The data that support the findings of this study are not publicly available but can be purchased at \url{https://lobsterdata.com}.

\subsection*{Code availability}
The code that reproduces this study for section 3 can be found at \url{https://sobigdata.d4science.org/catalogue-sobigdata?path=/dataset/score-driven_bayesian_online_change_point_detection_sd-bocpd_} or \url{https://gitlab.com/YvTsak/ScoreDrivenBOCPD} and for section 4 at \url{https://sobigdata.d4science.org/catalogue-sobigdata?path=/dataset/online_learning_of_order_flow_and_market_impact_olofmi_} or \url{https://gitlab.com/YvTsak/olofmi}.

\appendix

\section{Price dynamics inside an order flow regime}\label{appA}

\subsection{Concavity of order flow regimes}\label{appA1}
In table \ref{tab4} we report the number of regimes that satisfy the condition:
\begin{equation}\label{eq:condition of regimes}
   Z_R:=\left| \frac{\sum_{t_R\leq t< s_R}x_t}{\sum_{t_R\leq t< s_R}|x_t|}\right |>\Theta
\end{equation}
for various values of $\Theta$.  

\begin{center}
\begin{tabular}{c||c|c||c|c}
\hline
 & \multicolumn{2}{c||}{TSLA} & \multicolumn{2}{c}{MSFT} \\
\cline{1-5}
$\Theta$ & $1$ min & $3$ min & $1$ min & $3$ min \\
\hline\hline
0 & 911 & 546 & 1394 & 690 \\
0.5 & 674 & 466 & 1195 & 651 \\
0.9 & 344 & 321 & 827 & 550 \\  
  \hline
\end{tabular}
\captionof{table}{Number of regimes satisfying the condition in Eq. (\ref{eq:condition of regimes}) for $\Theta=0,0.5$ and $0.9$, for both TSLA and MSFT with $\Delta t=1$ min and $\Delta t = 3$ min.}
\label{tab4}
\end{center}

\subsection{Testing the square root impact law}
\label{AppA2}

 Table \ref{tab5}  presents the estimation of the parameters of the regression in Eq. (\ref{eq: regression}) when we consider the entire data sets without removing any outliers.  We observe that the exponent $\gamma$ is a bit larger than $1/2$ being typically close to $0.7$.

The outlier removal is obtained by the standard interquartile approach. Namely, we compute the first and third quartile $Q1$ and $Q3$, respectively, of log-returns. Then the data points outside the range $[Q1-1.5IQR,Q3+1.5IQR]$, where $IQR=Q3-Q1$,are considered as outliers.

\begin{figure}[!htb]
    \centering
    \subfigure[]{\includegraphics[width=0.44\textwidth]{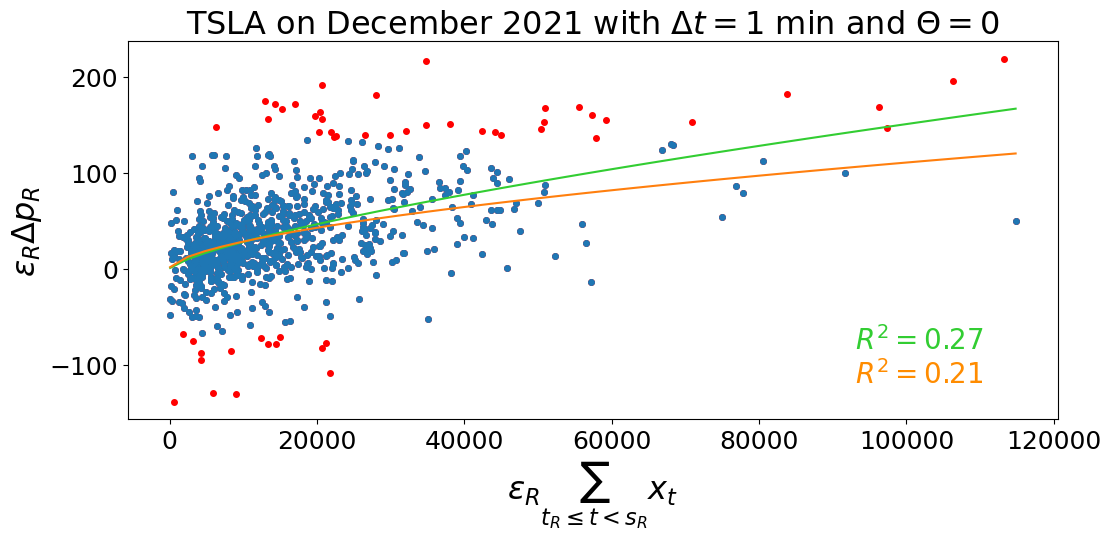}}
    \subfigure[]{\includegraphics[width=0.44\textwidth]{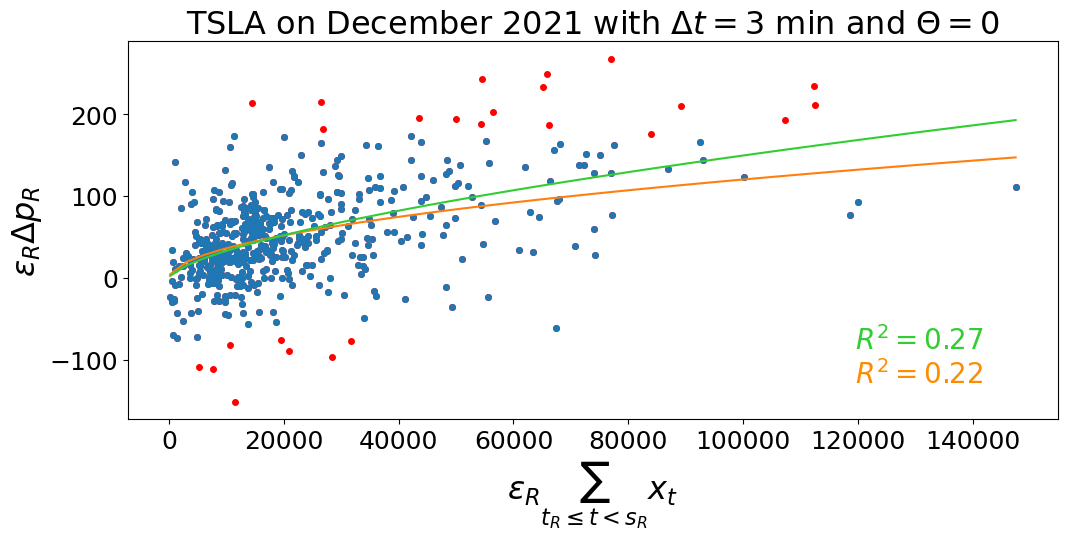}}  
    \subfigure[]{\includegraphics[width=0.44\textwidth]{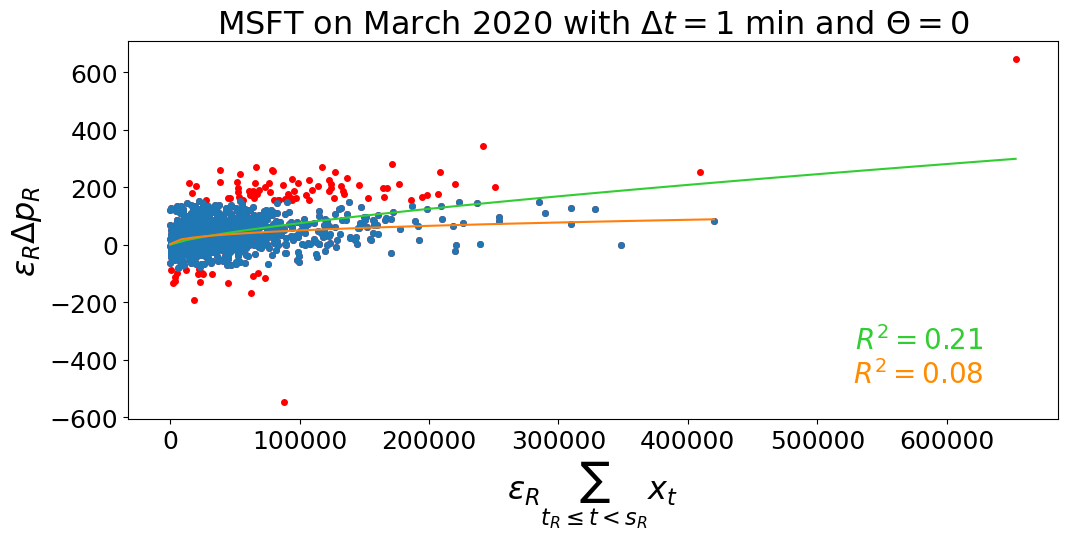}}
    \subfigure[]{\includegraphics[width=0.44\textwidth]{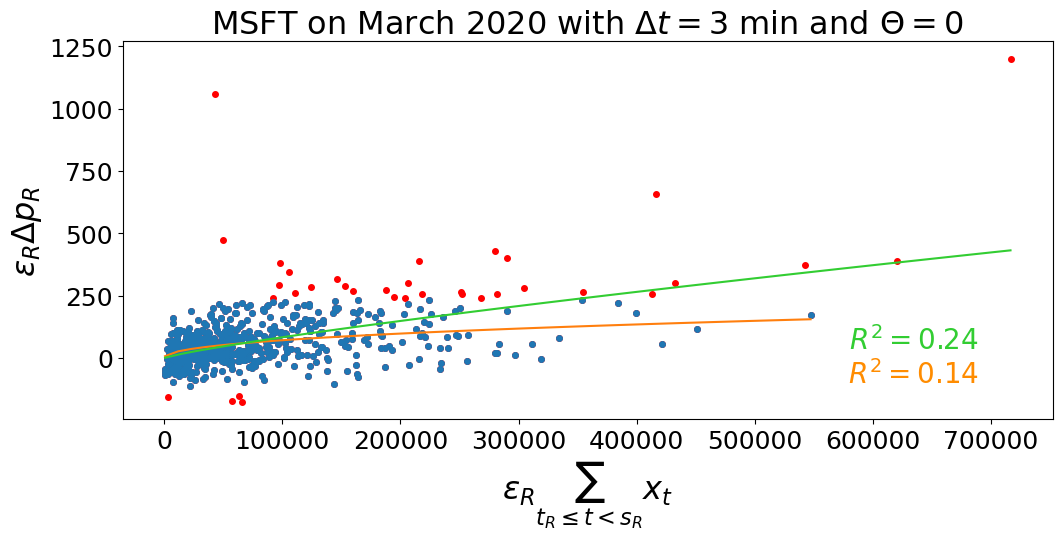}}  
    \caption{Data points and best fitting curves according to the regression of Eq. (\ref{eq: regression}) with $\Theta=0$ for Tesla (a) with $\Delta t=1$ min, (b) with $\Delta t=3$ min and for Microsoft  (c) with $\Delta t=1$ min, (d) with $\Delta t=3$ min. The plotted quantities are in basis points. The red points are the outliers. The green line is the fitting curve of the entire data set, while the orange line is the fitting curve when the outliers are excluded.}
    \label{fig8}
\end{figure}

Figure \ref{fig8} illustrates the data and the fits for the two stocks and the two timescales. The red points are those which are identified as outliers.

\begin{center}
\begin{tabular}{c||c c c c|c c c c}
\hline
& \multicolumn{8}{c}{TSLA}\\
\cline{2-9}
 & \multicolumn{4}{c|}{$1$ min} & \multicolumn{4}{c}{$3$ min}\\
$\Theta$ & $A$ & SE of $A$ & $\gamma$ & SE of $\gamma$ & $A$ & SE of $A$ & $\gamma$ & SE of $\gamma$ \\
\hline\hline
0 & $0.032$ & $0.013$ & $0.732$ & $0.04$ & $0.08$ & $0.04$ & $0.654$ & $0.047$\\

0.5 & $0.045$ & $0.018$ & $0.7$ & $0.039$ & $0.096$ & $0.047$ & $0.639$ & $0.046$\\

0.9 & $0.05$ & $0.023$ & $0.691$ & $0.044$ & $0.221$ & $0.101$ & $0.564$ & $0.043$\\  
  \hline
  \hline
& \multicolumn{8}{c}{MSFT}\\
\cline{2-9}
 & \multicolumn{4}{c|}{$1$ min} & \multicolumn{4}{c}{$3$ min}\\
$\Theta$ & $A$ & SE of $A$ & $\gamma$ & SE of $\gamma$ & $A$ & SE of $A$ & $\gamma$ & SE of $\gamma$ \\
\hline\hline
0 & $0.015$ & $0.006$ & $0.737$ & $0.037$ & $0.005$ & $0.003$ & $0.835$ & $0.053$\\

0.5 & $0.015$ & $0.006$ & $0.736$ & $0.038$ & $0.004$ & $0.003$ & $0.845$ & $0.055$\\

0.9 & $0.022$ & $0.01$ & $0.706$ & $0.042$ & $0.006$ & $0.004$ & $0.827$ & $0.061$\\  
  \hline
\end{tabular}
\captionof{table}{Estimated parameters and their standard errors (SE) of the regression of Eq. (\ref{eq: regression}). 
The whole dataset is considered.}
\label{tab5}
\end{center}

\section{Comparison of order flow  and market impact predictions under different regime shift models}\label{appB}
\begin{figure}[!htb]
    \centering
    \subfigure[]{\includegraphics[width=0.24\textwidth]{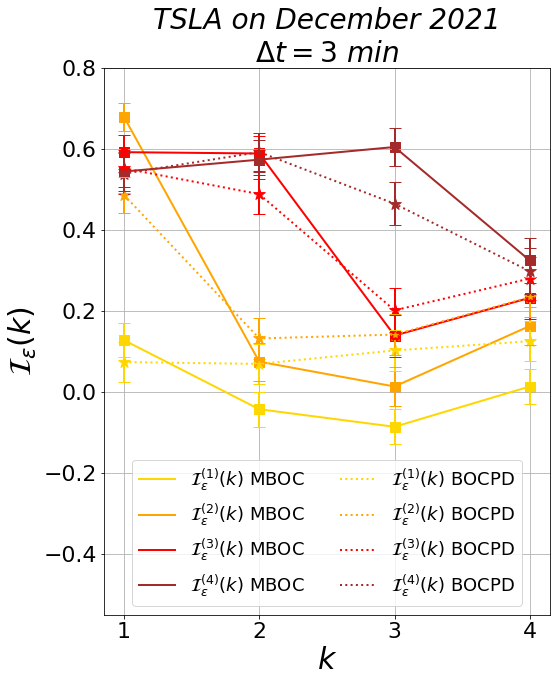}}
    \subfigure[]{\includegraphics[width=0.24\textwidth]{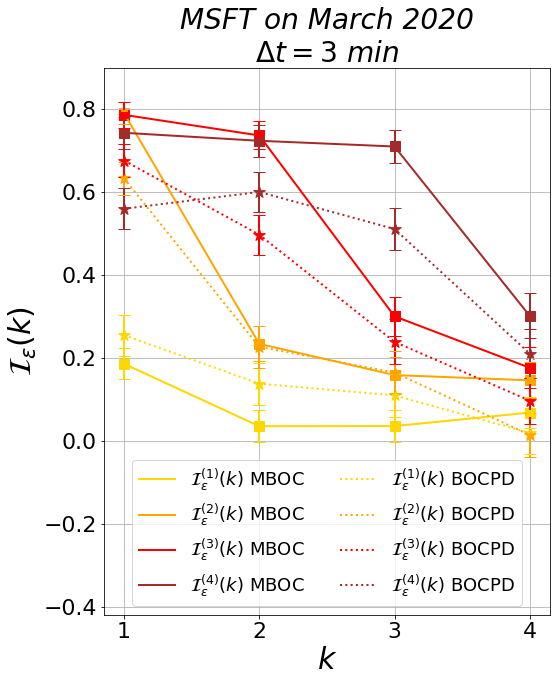}} 
    \caption{Online order flow prediction according to MBOC (solid lines) and BOCPD (dotted lines) models with $\Delta t = 3$ min (a) for Tesla and (b) for Microsoft.}
    \label{fig9}
\end{figure}

\begin{figure}[!htb]
    \centering
    \subfigure[]{\includegraphics[width=0.24\textwidth]{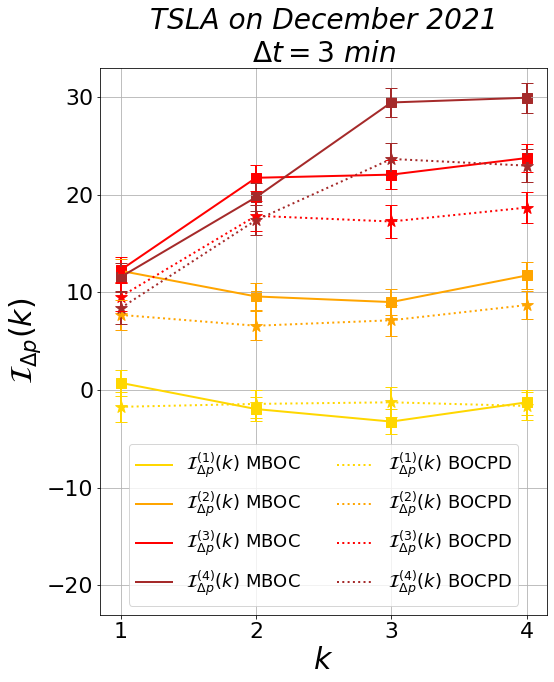}}  
    \subfigure[]{\includegraphics[width=0.24\textwidth]{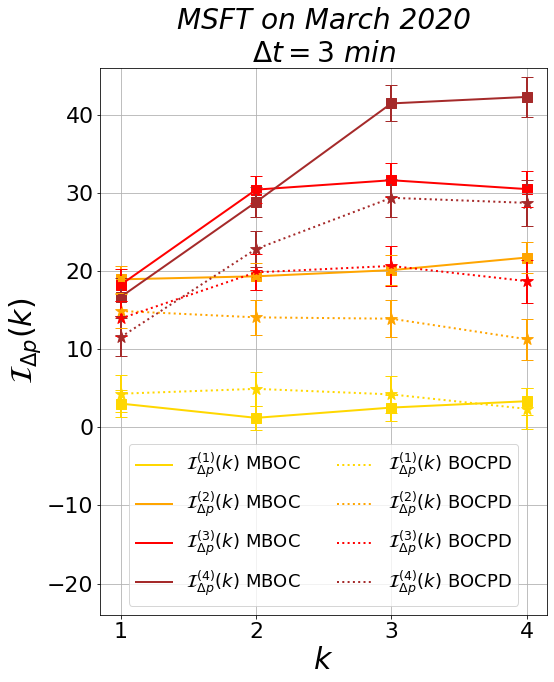}}  
    \caption{Online market impact prediction according to MBOC (solid lines) and BOCPD (dotted lines) models with $\Delta t = 3$ min (a) for Tesla  and (b) for Microsoft.}
    \label{fig10}
\end{figure}

Figure \ref{fig9} compares the correlation function $\mathcal{I}^{(m)}_\epsilon(k)$ of order flow for the MBOC and the BOCPD model. Figure \ref{fig10} compares the predictor of market impact $\mathcal{I}^{(m)}_{\Delta p}(k)$ for the same models.

From both figures, it is evident that the MBOC model outperforms the BOCPD when $m>1$. This justifies why in the main text we present the results obtained with the MBOC model.


\begin{thebibliography}{}

\bibitem[Adams and MacKay, 2007]{c:07} Adams, R. P., and MacKay, D. J. (2007). Bayesian online changepoint detection. arXiv preprint arXiv:0710.3742.

\bibitem[Almgren et al., 2005]{Almgren2005DirectEO} Almgren, R., Thum, C., Hauptmann, E., and Li, H. (2005). Direct estimation of equity market impact. Risk, 18(7), 58-62.

\bibitem[Bacry et al., 2015]{Bacry2015} Bacry, E., Iuga, A., Lasnier, M., and Lehalle, C.-A. (2015). Market impacts and the life cycle of investors orders. Market Microstructure and Liquidity, 1:1550009.

\bibitem[Barry and Hartigan, 1992]{r:92} Barry, D. and Hartigan, J. A. (1992). Product partition models for change point models. The Annals of Statistics, 20:260 – 279.

\bibitem[Bershova and Rakhlin, 2013]{ar:2013} Bershova, N. and Rakhlin, D. (2013). The non-linear market impact of large trades: evidence from buy-side order flow. Quantitative Finance, 13:1759–1778.

\bibitem[Blasques et al., 2014]{Score-Driven_AR} Blasques, F., Koopman, S., and Lucas, A. (2014). Optimal formulations for nonlinear autoregressive processes. WorkingPaper 14-103/III, Tinbergen Institute.

\bibitem[Bouchaud et al., 2018]{book:18} Bouchaud, J.-P., Bonart, J., Donier, J., and Gould, M. (2018). Trades, Quotes and Prices: Financial Markets Under the Microscope. Cambridge University Press.

\bibitem[Bouchaud et al., 2009]{em:09} Bouchaud, J.-P., Farmer, J. D., and Lillo, F. (2009). How markets slowly digest changes in supply and demand. Handbook of Financial Markets: Dynamics and Evolution, Handbook of Finance.

\bibitem[Bouchaud et al., 2004]{Bouchaud04} Bouchaud, J.-P., Gefen, Y., Potters, M., and Wyart, M. (2004). Fluctuations and response in financial markets: the subtle nature of ‘random’ price changes. Quantitative Finance, 4:176–190.

\bibitem[Bouchaud et al., 2006]{BouchaudKockelkorenPotters}
J.P. Bouchaud, J. Kockelkoren and M. Potters (2006).
Random walks, liquidity
molasses and critical response in financial markets. Quantitative Finance, 6:115-123.

\bibitem[Cox, 1981]{cox} Cox, D. (1981). Statistical analysis of time series: Some recent developments. Scandinavian Journal of Statistics, 8:93 – 115.

\bibitem[Creal et al., 2013]{Score-Driven1} Creal, D., Koopman, S. J., and Lucas, A. (2013). Generalized autoregressive score models with applications. Journal of Applied Econometrics, 28:777–795.

\bibitem[Diaconis and Ylvisaker, 1979]{r:79} Diaconis, P. and Ylvisaker, D. (1979). Conjugate priors for exponential families. The Annals of Statistics, 7:269–281.

\bibitem[Diebold and Inoue, 2001]{r:01} Diebold, F. X. and Inoue, A. (2001). Long memory and regime switching. Journal of Econometrics, 105:131–159.

\bibitem[Donier et al., 2015]{LLOB} Donier, J., Bonart, J., Mastromatteo, I., and Bouchaud, J. P. (2015). A fully consistent, minimal model for non-linear market impact. Quantitative Finance, 15:1109 – 1121.

\bibitem[Fan and Mackey, 2017]{r:2017} Fan, Z. and Mackey, L. (2017). Empirical Bayesian analysis of simultaneous changepoints in multiple data sequences. The Annals of Applied Statistics, 11:2200–2221.

\bibitem[Fearnhead and Liu, 2007]{r:07} Fearnhead, P. and Liu, Z. (2007). On-line inference for multi- ple changepoint problems. Journal of the Royal Statistical Society, 69:589–605.

\bibitem[Ghahramani, 2015]{r:15} Ghahramani, Z. (2015). Probabilistic machine learning and artificial intelligence. Nature, 521:452–459.

\bibitem[Granger and Hyung, 1999]{GrangerHyung}
C.W.J. Granger and N. Hyung (1999).
Occasional  structural  breaks  and  long  memory.
Department of Economics, University of California, San Diego.

\bibitem[Harvey, 2013]{Harvey} Harvey, A. (2013). Dynamic models for volatility and heavy tails: with applications to financial and economic time series. Cambridge University Press.

\bibitem[Lillo, 2023]{Lillo23} Lillo, F. (2023). Order flow and price formation, volume Machine Learning and Data Sciences for Financial Markets: A Guide to Contemporary Practices. Cambridge University Press.

\bibitem[Lillo and Farmer, 2004]{LilloFarmer04} Lillo, F. and Farmer, J. D. (2004). The long memory of efficient market. Studies in Nonlinear Dynamics and Econometrics, 8:1.

\bibitem[Lillo et al., 2005]{ar:2005} Lillo, F., Mike, S., and Farmer, J. (2005). Theory for long memory in supply and demand. Physical Review E, 71:066122.

\bibitem[Lleo et al., 2022]{r:22} Lleo, S., Zhitlukhin, M., and Ziemba, W. T. (2022). Using a mean-changing stochastic processes exit–entry model for stock market long-short prediction. The Journal of Portfolio Management, 49:172–197.

\bibitem[Lleo et al., 2023]{LleoZiembaLi}
Sébastien Lleo, William T. Ziemba and Jessica Li (2023).
Do Factor Models Explain Breaks in the Distribution of Equity Returns?
The Journal of Portfolio Management

\bibitem[Lleo et al., 2020]{c:2020} Lleo, S., Ziemba, W. T., and Li, J. (2020). Exploring breaks in the distribution of stock returns: Empirical evidence from apple inc. In SSRN Working Paper 3700419. Elsevier.

\bibitem[Mikosch and Starica, 1999]{MikoschStarica}
C. Mikosch and C. Starica (1999).
Change of structure in financial time series, long range dependence and the GARCH model. Technical Report, preprint available at http,//www.cs.nl/ eke/iwi/preprints.

\bibitem[Moro et al., 2009]{r:09} Moro, E., Vicente, J., Moyano, L. G., Gerig, A., Farmer, J. D., Vaglica, G., Lillo, F., and Mantegna, R. N. (2009). Market impact and trading profile of hidden orders in stock markets. Physical Review E, 80:452–459.

\bibitem[Murphy, 2007]{k:07} Murphy, K. P. (2007). Conjugate Bayesian analysis of the Gaussian distribution. Technical report, University of British Columbia.

\bibitem[Patzelt and Bouchaud, 2018]{PhysRevE.97.012304} Patzelt, F. and Bouchaud, J.-P. (2018). Universal scaling and nonlinearity of aggregate price impact in financial markets. Physical Review E, 97:012304.

\bibitem[Sato and Kanazawa, 2023a]{c:23} Sato, Y. and Kanazawa, K. (2023a). Can we infer microscopic financial information from the long memory in market-order flow?: a quantitative test of the Lillo-Mike-Farmer model. arXiv:2301.13505.

\bibitem[Sato and Kanazawa, 2023b]{kana23} Sato, Y. and Kanazawa, K. (2023b). Exact solution to a generalised lillo-mike-farmer model with heterogeneous order-splitting strategies. arXiv:2306.13378.

\bibitem[Taranto et al., 2018]{Taranto-TothA}
D.E. Taranto and G. Bormetti and J.P. Bouchaud and F. Lillo and B. Toth (2018).
Linear models for the impact of order flow on prices. I. History dependent impact models. Quantitative Finance, 18:903-915.

\bibitem[Taranto et al., 2018]{Taranto-TothB}
D.E. Taranto and G. Bormetti and J.P. Bouchaud and F. Lillo and B. Toth (2018).
Linear models for the impact of order flow on prices. II. The Mixture Transition Distribution model. Quantitative Finance, 18:917-931.

\bibitem[Torre, 1997]{BARRA} Torre, N. (1997). Barra market impact model handbook. (BARRA Inc., Berkeley, 1997).

\bibitem[To\'th et al., 2011]{PhysRevX.1.021006} T\'oth, B. and Lemp\'eri\`ere, Y. and Deremble, C. and de Lataillade, J. and Kockelkoren, J. and Bouchaud, J.-P. (2011). Anomalous price impact and the critical nature of liquidity in financial markets. Physical Review X, 1:021006.

\bibitem[T\'oth et al., 2015]{Toth} T\'oth, B., Palit, I., Lillo, F., and Farmer, J. D. (2015). Why is equity order flow so persistent? Journal of Economic Dynamics and Control, 51:218–239.

\bibitem[To\'th et al., 2016]{SRILforOptions} To\'th, B., Eisler, Z., and Bouchaud, J.-P. (2016). The square-root impact law also holds for option markets. Wilmott, 85.

\bibitem[Tsaknaki et al., 2023]{tsaknaki} Tsaknaki, I.-Y., Lillo, F., and Mazzarisi, P. (2023). A score-driven Bayesian online change-point detection model. (in preparation).

\bibitem[Vaglica et al., 2010]{r:80} Vaglica, G., Lillo, F., and Mantegna, R. N. (2010). Statistical identification with hidden Markov models of large order splitting strategies in an equity market. New Journal of Physics, 12:075031.

\bibitem[Vaglica et al., 2008]{r:20} Vaglica, G., Lillo, F., Moro, E., and Mantegna, R. N. (2008). Scaling laws of strategic behavior and size heterogeneity in agent dynamics. Physical Review E, 77:036110.

\bibitem[Wainwright and Jordan, 2008]{r:08} Wainwright, M. J. and Jordan, M. I. (2008). Graphical models, exponential families, and variational inference. Foundations and Trends in Machine Learning, 1:1–305.

\bibitem[Xuan and Murphy, 2007]{c:07b} Xuan, X. and Murphy, K. (2007). Modeling changing dependency structure in multivariate time series. In Proceedings of the international conference on Machine learning (ICML-07), volume 24, pages 1055–1062. PMLR.

\bibitem[Yamamoto and Lebaron, 2010]{YamamotoLebaron}
R. Yamamoto and B. Lebaron (2010).
Order-splitting and long-memory in an order-driven market. The European Physical Journal B-Condensed Matter and
Complex Systems, 73:51-57.

\bibitem[Zarinelli et al., 2015]{r:2015} Zarinelli, E., Treccani, M., Farmer, J., and Lillo, F. (2015). Beyond the square root: Evidence for logarithmic dependence of market impact on size and participation rate. Market Microstructure and Liquidity, 1:1550004.

\bibitem[Zhao et al., 2022]{c:2022} Zhao, Y., Landgrebe, E., Shekhtman, E., and Udell, M. (2022). Online missing value imputation and change point detection with the Gaussian copula. In Proceedings of the AAAI Conference on Artificial Intelligence (AAAI-22), volume 36, pages 9199–9207.

\end{thebibliography}

\end{document}